\documentclass[useAMS,usenatbib]{mn2e}
\usepackage{graphicx} 
\usepackage{times}
\usepackage{epsfig}
\usepackage{epstopdf}
\usepackage{amsfonts}
\usepackage{amsmath}
\usepackage{amsbsy}
\usepackage{bm}
\usepackage{url}
\usepackage{microtype}
\usepackage{rotating}
\usepackage{sidecap}
\usepackage{longtable}
\usepackage{lscape}
\usepackage{rotating}
\usepackage{color}
\usepackage{amssymb,xcolor}
\usepackage[normalem]{ulem}

\def\ergs{{\rm erg\,\,s^{-1}}}
\def\cm2{{\rm cm^{-2}}}
\def\ergcms{{\rm erg\,cm^{-2}\,s^{-1}}}
\def\NH{N_\textrm{H}}
\def\cts{{\rm counts\,s^{-1}}}
\newcommand{\swift}{\textsl{Swift}}
\newcommand{\inte}{\textsl{INTEGRAL}}

\title[Disk outflows from V404 Cyg]{\textit{Swift} observations of V404 Cyg during the 2015 outburst: \\
X-ray outflows from super-Eddington accretion}

\author[S. Motta et al.]{S.E. Motta$^{1}$, J. J. E. Kajava$^{2,3,4}$, C. S\'anchez-Fern\'andez$^4$, A.P. Beardmore$^{5}$, A. Sanna$^{6}$,  
\and K.L. Page$^{5}$, R. Fender$^{1}$, D. Altamirano$^{7}$, P. Charles$^{1,7,8}$, M. Giustini$^{9}$, C. Knigge$^{7}$, E. \and Kuulkers$^{4,10}$, S. Oates$^{11,12}$, J.P. Osborne$^{5}$\\
$^{1}$University of Oxford, Department of Physics, Astrophysics, Denys Wilkinson Building, Keble Road, Oxford OX1 3RH, UK\\
$^{2}$Finnish Centre for Astronomy with ESO (FINCA), University of Turku, V\"{a}is\"{a}l\"{a}ntie 20, FIN-21500 Piikki\"{o}, Finland\\
$^{3}$Tuorla Observatory, University of Turku, V\"{a}is\"{a}l\"{a}ntie 20, FIN-21500 Piikki\"{o}, Finland\\
$^{4}$European Space Astronomy Centre (ESA/ESAC), Science Operations Department, 28691 Villanueva de la Ca\~{n}ada, Madrid, Spain\\
$^{5}$Dept. of Physics and Astronomy, University of Leicester, Leicester, LE1 7RH, UK\\
$^{6}$Dipartimento di Fisica, Universit\`a degli Studi di Cagliari, SP Monserrato-Sestu km 0.7, 09042 Monserrato, Italy\\
$^{7}$Dept. of Physics \& Astronomy, University of Southampton, Southampton, Hampshire SO17 1BJ, UK\\
$^{8}$Instituto de Astrof\'isica de Canarias, E-38205 La Laguna, S/C de Tenerife, Spain\\
$^{9}$SRON, Netherlands Institute for Space Research, Sorbonnelaan 2, 3584 CA Utrecht, The Netherlands\\
$^{10}$ESA/ESTEC, Keplerlaan 1, 2201 AZ Noordwijk, The Netherlands\\
$^{11}$Department of Physics, University of Warwick, Coventry, CV4 7AL, UK\\
$^{12}$Instituto de Astrof\'isica de Andaluc\'ia (IAA-CSIC), Glorieta de la Astronom\'ia s/n, E-18008, Granada, Spain\\
}
\begin{document}
\maketitle

\begin{abstract}

\noindent 
The black-hole binary V404 Cyg entered the outburst phase in June 2015 after 26 years of X-ray quiescence, and with its behaviour broke the outburst evolution pattern typical of most black-hole binaries.
We observed the entire outburst with the \textit{Swift} satellite and performed time-resolved spectroscopy of its most active phase, obtaining over a thousand spectra with exposures from tens to hundreds of seconds.
All the spectra can be fitted with an absorbed power law model, which most of the time required the presence of a partial covering. A blue-shifted iron-K$\alpha$ line appears in 10\% of the spectra together with the signature of high column densities, and about 20\% of the spectra seem to show signatures of reflection. None of the  spectra showed the unambiguous presence of soft disk-blackbody emission, while the observed bolometric flux exceeded the Eddington value in 3\% of the spectra.
Our  results can be explained assuming that the inner part of the accretion flow is inflated into a \textit{slim disk} that both hides the innermost (and brightest) regions of the flow, and produces a cold, clumpy, high-density outflow that introduces the high-absorption and fast spectral variability observed. We argue that the black hole in V404 Cyg might have been accreting erratically or even continuously at Eddington/Super-Eddington rates -- thus sustaining a surrounding slim disk -- while being partly or completely obscured by the inflated disk and its outflow. Hence, the largest flares produced by the source might not be accretion-driven events, but instead the effects of the unveiling of the extremely bright source hidden within the system.
\end{abstract}



\section{Introduction}

Black hole (BH) X-ray binaries (BHBs) are typically transient systems, which alternate between relatively short outbursts and long periods of (X-ray) quiescence. During outbursts most BHBs show significant luminosity changes from $L \sim 10^{30-31}\,\ergs$ in quiescence (see \citealt{Wijnands2015}) to $L \sim 10^{38-39}\,\ergs$ or more in outburst, and display a ``hysteresis'' behaviour that becomes apparent as q-shaped loops in the so-called Hardness-Intensity diagram (HID; see e.g., \citealt{Homan2001}). These cyclic patterns have a clear and repeatable association with mechanical feedback in the form of winds and relativistic jets (see \citealt{Fender2009} and \citealt{Ponti2012}).

The hard state (low-hard state, LHS, see e.g. \citealt{Belloni2016} and references therein, \citealt{Homan2001}, \citealt{McClintock2006}) occupies the right-hand regions of the HID, where the source X-ray spectrum is dominated by emission from Compton up-scattering of soft seed photons either produced in a cool geometrically thin accretion disk truncated at large radii, by synchrotron-self-Compton emission from hot electrons located close to the central black hole (e.g., \citealt{Poutanen2014}), or by synchrotron emission from a compact jet (e.g. \citealt{Markoff2010}). The soft state (high-soft state, HSS, e.g. \citealt{Belloni2016}) corresponds to the left-hand side of the HID and the emission of the source is typically dominated by thermal emission from a geometrically thin accretion disk extending down to the innermost stable circular orbit around the BH (\citealt{Shakura1973}).
In between these two states lie the so-called \textit{intermediate} states (hard and soft intermediate states, HIMS and SIMS, respectively), where the energy spectra show the properties of both the LHS and the HSS evolving smoothly during the transitions between the LHS and HSS.
An iron K-$\alpha$ emission line at $\sim$ 6.4 keV is often detected in the hard states, where it is accompanied by a modest reflection hump (resulting from the non-thermal emission of a Comptonizing medium reflecting off the disk) peaking at energies around 30 keV \citep{Reynolds2014}. As BHBs evolve to the soft states, the intensity of the reflection features (i.e. line and reflection hump) increases (e.g., \citealt{Zdziarski1999, Plant2014}), with a simultaneous blurring/broadening of the iron K-$\alpha$ line that -- believed to be produced close to the central BH -- is significantly changed by relativistic effects. 

\subsection{V404 Cygni}

V404 Cyg was first identified as an optical nova in 1938 and later associated with the X-ray transient GS~2023+338, discovered by \textit{Ginga} at the beginning of its X-ray outburst in 1989 (\citealt{Makino1989}). The 1989 outburst displayed extreme variability and V404~Cyg became temporarily one of the brightest sources ever observed in X-rays. 
\citet{Casares1992} determined the orbital period of the system ($\sim$6.5 days) and \cite{Miller-Jones2009} determined the distance to the source through radio parallax ($d = 2.39 \pm 0.14$~kpc). 
\cite{Casares1992} also obtained the first determination of the system's mass function ($f$(M) = 6.26 $\pm$ 0.31 M$_{\odot}$), confirming the black hole nature of the compact object in V404 Cyg and classifying it as a low-mass X-ray binary.

V404 Cyg seems to break the typical BHB pattern, showing a dynamical range spanning more than 3 orders of magnitude in flux on time-scales as short as a few minutes or even seconds (as opposed to the typical long time-scale evolution of a standard BHB in outburst), and spectral variability that almost never resembled that usually seen in other BHBs. Additionally, the long term variability of the source is severely affected by heavy absorption (\citealt{Oosterbroek1996}, \citealt{Zycki1999}). Hence, as expected, the V404 Cyg HID does not seem to show the typical hysteresis loops observed in other BHBs.

Already during the 1989 outburst the source showed luminosities exceeding the Eddington limit (e.g. \citealt{Oosterbroek1996}), but without showing a canonical disk-dominated HSS (however, see \citealt{Zycki1999}, who reported on a short-lived disk-dominated state). The 1989 outburst was characterized by extreme variability, possibly partly due to accretion events (somewhat similar to those seen in GRS 1915+105, see \citealt{Belloni1997a}), but also ascribed to a heavy and very variable photo-electric local absorption (\citealt{Oosterbroek1996}, \citealt{Zycki1999}). Despite its peculiarity, however, V404 Cyg was among one of the first sources where the radio-X-ray correlation typical of accreting systems was found, and it appeared to be consistent with that shown by well-behaved systems such as GX 339-4 (e.g., \citealt{Gallo2003}). It is also noteworthy that the quiescent X-ray emission of V404 Cyg is characterized by an X-ray luminosity ($L_\textrm{x} \sim 10^{34}\,\ergcms$) significantly higher than that of other BHBs in quiescence, and by a well-measured flat radio spectrum. This led to the establishment of the presence of a compact steady jet in the quiescence state of the source  (\citealt{Gallo2005}).

\bigskip

On 2015 June 15 18:32 UT (MJD 57188.772), the \swift/BAT triggered on a bright hard X-ray flare from a source that was soon recognized to be the BH low mass X-ray binary V404 Cyg  back in outburst after 26 years of quiescence (\citealt{Barthelmy2015}, \citealt{Kuulkers2015}).
V404 Cyg reached the outburst peak on June 26 and then began a rapid fading towards X-ray quiescence, that was reached between 2015 August 5 and August 21 (\citealt{Sivakoff2015}). Throughout this outburst the source displayed highly variable multi-wavelength activity (e.g., \citealt{Rodriguez2015}, \citealt{Gandhi2016}, \citealt{Kimura2016}, \citealt{Bernardini2016}), that was monitored by the astronomical community through one of the most extensive observing campaigns ever performed on an X-ray binary outburst (see \citealt{Sivakoff2015}, and references therein).

\textit{INTEGRAL}/IBIS-ISGRI observations of the system during the June 2015 outburst suggest that the X-ray spectrum of V404 Cyg is not too different from the spectrum of a standard BHB  when the effects of heavy absorption are accounted for (i.e. at energies above 25\,keV; \citealt{Sanchez-Fernandez2017}), even though during the June 2015 outburst the system most likely never reached a proper disk-dominated HSS (see also \citealt{Motta2017}).
Thus, it remains clear that V404 Cyg retains some very peculiar characteristics that are not typically seen in other BHBs. 

\cite{Munoz-Darias2016} reported the detection of clear P-Cyg profiles in the system's optical spectra over the entire active phase, which suggest the presence of a sustained outer accretion disc wind in V404 Cyg, never observed before at optical wavelengths in other BHBs. 
The outflowing wind was neutral,  expanded at 1\% of the speed of light and had a large covering fraction.  Such an outflow triggered a nebular phase once accretion sharply dropped and the ejecta became optically thin. 
\cite{King2015} also reported the presence, during the highest flux phases of the outburst, of P-Cyg profiles associated with the several X-ray lines observed in the \textit{Chandra} spectrum of V404 Cyg, consistent with being the effect of a strong disk wind, likely radiation or thermally driven as the source approached (or even exceeded) its Eddington limit. 

The presence of (ionized and/or neutral) outflows and ``envelopes'' seems to be a characteristic of a number of sources\footnote{Note, however, that the well-known super-Eddington accretor GRS 1915+105 does not show any trace of high-column density envelopes/outflows, while it does show ionized winds \citep{Neilsen2009}.} accreting matter at a rate comparable with or above the Eddington limit. An example is the BH X-ray transient V4641 Sgr (\citealt{Revnivtsev2002}), which in 1999 showed an episode of super-Eddington accretion onto the BH. During the outburst an extended optically thick envelope/outflow was formed around the source. During this event V4641 Sgr appeared in many aspects very similar to SS 433 \citep{Margon1984}, known for the presence of obscuring material shielding the central source from the observer. Another case is represented by the BHB GRO J1655-40, which during both the outbursts observed by RXTE in 1995 and 2005 entered the so-called Ultra-luminous state (see \citealt{Motta2012}), when high luminosity flares were accompanied by slight increments in  the local column density (e.g. \citealt{Kalemci2016}). Later on GRO J1655-40 entered a so-called ``hyper-soft'' state  (\citealt{Uttley2015}) during which it showed signatures of variable absorption and scattering due to a Compton-thick, fully ionized disk wind (see also \citealt{Neilsen2016} and \citealt{Shidatsu2016}).

Here we report the results of a detailed study of the soft X-ray spectra of V404 Cyg  as observed by \textit{Swift}/XRT, during the most active phase of the June 2015 outburst. Such a phase roughly corresponds to the first two weeks of activity after the outburst start, when V404 Cyg showed intense flaring alternating with phases of very low flux, low variability emission. We focus in particular on the variations of the spectral properties induced by the presence of a variable neutral absorber produced by the source itself (see \citealt{Motta2017}) in the inner regions of the accretion flow, where most of the X-ray emission arises from. Results from the analysis of \textit{INTEGRAL}/IBIS-ISGRI data and from a small \textit{INTEGRAL}/JEM-X, \textit{INTEGRAL}/IBIS-ISGRI and \textit{Swift}/XRT simultaneous dataset have been reported by \citet{Sanchez-Fernandez2017} and \cite{Motta2017}, respectively.

\begin{figure}
\centering
\includegraphics[width=0.45\textwidth]{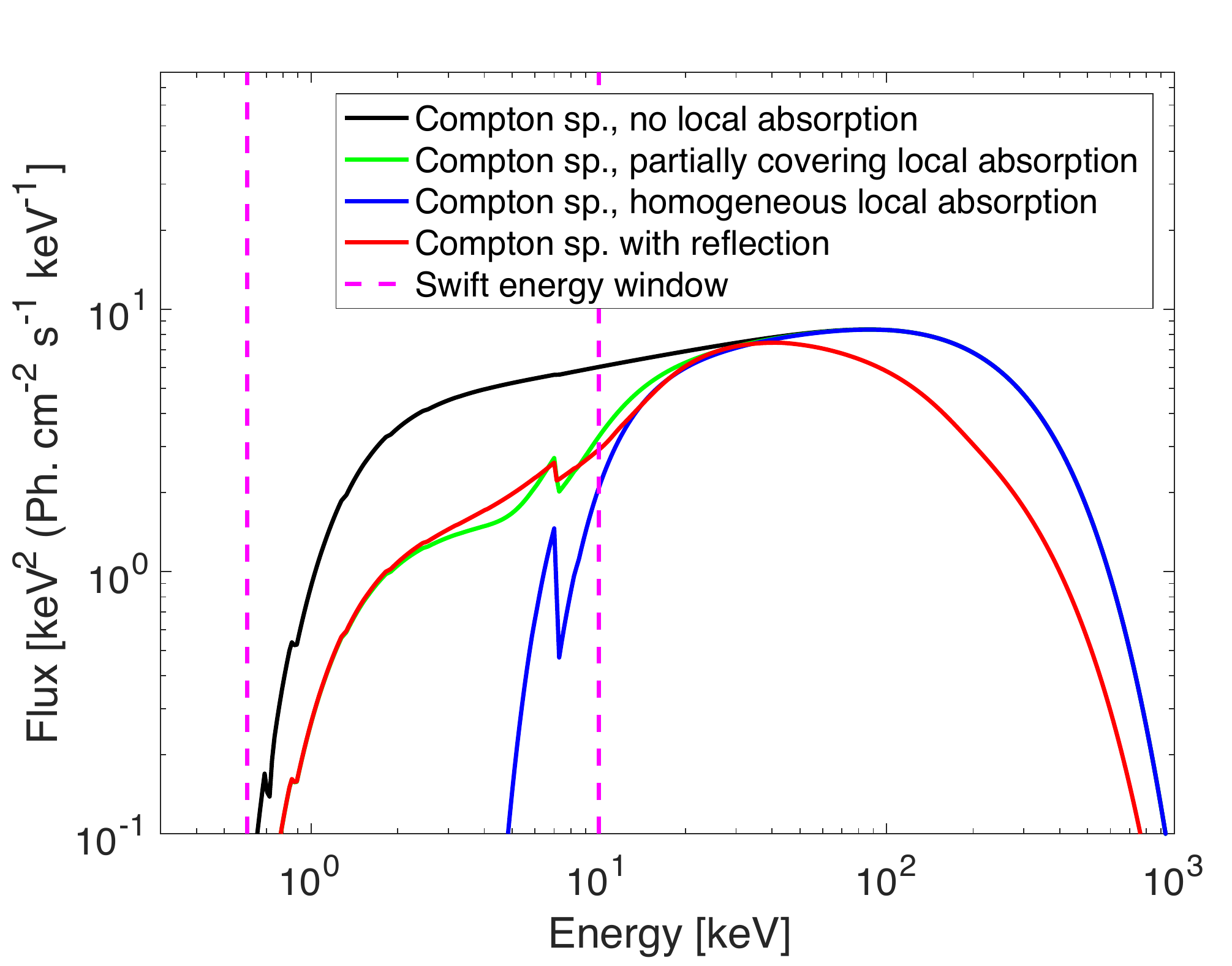}
\caption{Examples of different models to illustrate the spectral variability encountered. The spectral continuum used in this figure is a Compton spectrum (\textsc{compps}), adopted to show the high-energy behaviour of the spectra (not seen in \textit{Swift} data). The magenta dashed lines delimit the \textit{Swift} energy range. For all the spectra we set the electron temperature to 60 keV and seed photon temperature to 0.1 keV and we varied the properties of the absorber and of the reflector. 
Black line: Compton spectrum modified by interstellar absorption only. Orange line: Compton spectrum modified by interstellar absorption and a partially (85\%) covering absorber ($N_\textrm{H} = 10^{24}\,\cm2$). Blue line: Compton spectrum modified by interstellar absorption and by a homogeneous (100\%) covering absorber ($N_\textrm{H} = 10^{24}\,\cm2$). Red line: Compton spectrum convolved with a reflection component (reflection fraction = 10), and modified by interstellar absorption. From the figure it is clear that it might be challenging to distinguish between a partially-covered absorbed spectrum and a reflected spectrum. A colour version of this Figure is available on-line.}\label{fig:modello}
\end{figure}

\begin{figure*}
\centering
\includegraphics[width=1.0\textwidth]{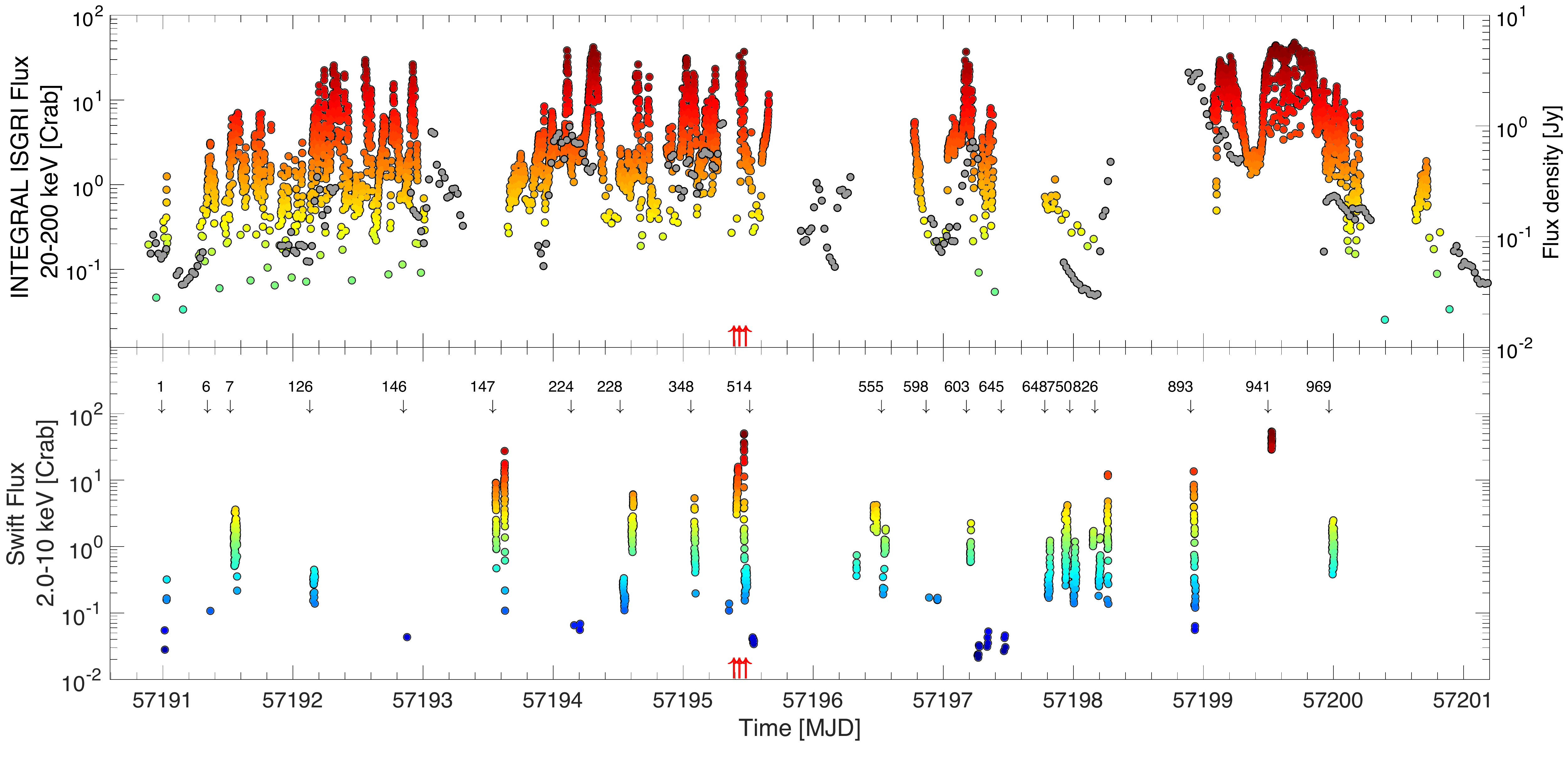}
\caption{\textit{Top panel}: \textit{INTEGRAL}/IBIS-ISGRI light-curve of the flaring phase of the June-August 2015 outburst of V404 Cyg between 20 and 200\,keV, and \textit{AMI} radio light curve (grey points) in Jy units (see right-hand side y-axis), extracted with a 0.02 days time resolution in a narrow frequency band centred at 14.6 GHz (Fender et al. in prep). 
The count rate has been converted to Crab units using the Crab count rates obtained during \textit{INTEGRAL} revolutions 1598 and 1599. 
\textit{Bottom panel:} \textit{Swift}/XRT light-curve of the flaring phase of the 2015 June-August outburst. Arrows indicate the Spectrum number we used to plot the fit parameters in Fig. \ref{fig:parameters}. In both panels, the points are colour-coded according to the flux, with redder points corresponding to higher fluxes.
Note that the flux scale is different in the two panels due to the different energy bands of \textit{INTEGRAL}/IBIS-ISGRI and \textit{Swift}/XRT. The colour-coding used in the bottom panel is the same used in Fig. \ref{fig:parameters} and Fig. \ref{fig:parameters_refl}. In both panels, the red arrows mark the eight ejection times reported by Tetarenko et al. 2017, based on Very Large Array, the Sub-millimeter Array and the James Clerk Maxwell Telescope data taken on MJD 57195 during four hours of observations. Two of these ejections occurred during our \swift/\textit{XRT} data, in particular in correspondence to spectra 470 and 484. Note that the eight ejections almost completely overlap to one another in our plot given their very short separation in time.
We refer the reader to Tetarenko et al. 2017 for a zoomed plot showing the ejection times with respect to the \textit{INTEGRAL}/IBIS-ISGRI light curve.
}
\label{fig:ISGRI}
\end{figure*}


\section{Observations and data analysis}\label{sec:observations} 

After the initial \swift/BAT trigger on 2015 June 15 (MJD~57188) \swift/XRT provided a total of 95 observations of V404 Cyg (often more than one per day), each containing one or more snapshots\footnote{An observation is a collection of consecutive snapshots of a single target that does not generally span more than one day, while a snapshot is an unbroken pointing in a given direction.} of variable duration, from the start of the outburst all the way down to quiescence.
The overall \swift/XRT exposure time was 171~ks from observations normally formed by a number of snapshots, taken in both Windowed Timing (WT) mode and Photon Counting (PC) mode \citep{Burrows2005}. 
The WT mode provides 1-dimensional imaging with a 1.7 milli-second time resolution, and allows bright sources to be observed without being affected by pile-up (up to relatively high count rates). The PC mode, instead, provides 2-dimensional imaging with a 2.5 second time resolution, and allows sources to be observed with count rates up to 1 $\cts$. Both the WT and PC modes provide data in the 0.3--10.0~keV energy band.

In this paper we focus on the most active phase of the source, which occurred during the first two weeks of the outburst. Thus we only considered observations performed in WT mode (see Table \ref{tab:log} and Fig. \ref{fig:ISGRI}), where the count rate in the source region is  $\geq 1\,\cts$ (and often hundreds of $\cts$),  which allowed us to obtain good S/N spectra. 

\cite{Beardmore2015} reported the discovery of a variable X-ray dust scattering halo peaking at 1--2 keV in the \textit{Swift}/XRT data (first seen on MJD 57192). The halo was bright enough to completely dominate the overall X-ray emission in the V404 Cyg field when the source was decaying to quiescence \citep{Vasilopoulos2016,Heinz2016}, but was present and at times dominant during a large fraction of the most active phase of the outburst (\citealt{Beardmore2016}). 
For this reason, among the WT observations, we only analysed data taken before June 27, 2015 (MJD 57200). 
After this date the halo started to dominate completely the field of view,\footnote{Note that the dust scattering halo in X-rays was filling completely the \textit{Swift}/XRT field of view.} while the source, already fainter than 1 $\cts$, slowly decayed to quiescence. 

\begin{table*}	
\centering									
\caption{\textit{Swift}/XRT observations considered in this work. All the observations listed were taken in \textit{Windowed Timing mode}. For each observation, we indicate the number of the first spectrum corresponding to each snapshot and the outer radius of the extraction region used for each snapshot (see Appendix \ref{App:analysis}, sec. \ref{sec:datared} for details). Note that sometimes snapshots from the same observations are spaced out by snapshots from another observation. The third snapshot of Obs. 00031403049 contained a large flare, followed by a very low count rate phase. Therefore, we extracted events in a region with an outer radius of 30 pixels during the flare, and 10 pixels during the low count rate phase, which was affected by the presence of the dust scattering halo. The second snapshot of Obs. 00033832001 showed high dust scattering halo contamination, therefore the spectrum had to be extracted in a 10-pixels radius region, despite the high source flux. However, the resulting pile-up corrected spectrum did not contain enough counts to be considered. 
}  
\begin{tabular}{c c c c c c c}														
\hline \hline			
Observation ID	&	Snapshots	&	Start time	&	Duration	& First spectrum	& Extraction region outer radius \\
				&	per Obs.	&	(UTC time)	&	(s)			& for each snapshot	& (pixels)							\\
\hline		\hline				
00031403038	&	2	&	2015-06-18 00:18:59	&	744	&	1, 6							&	20, 20						\\
00644520000	&	1	&	2015-06-18 00:28:19	&	4055	&	3							&	20							\\
00644627000	&	1	&	2015-06-18 12:53:33	&	2331	&	7							&	20							\\
00031403042	&	1	&	2015-06-19 03:44:59	&	1264	&	126							&	20							\\
00031403043	&	1	&	2015-06-19 21:00:59	&	1490	&	146							&	10							\\
00031403040	&	2	&	2015-06-20 13:25:20	&	1449	&	147, 181					&	20, 20						\\
00031403048	&	2	&	2015-06-21 03:55:18	&	994		&	224, 225					&	10, 10						\\
00031403046	&	2	&	2015-06-21 12:54:59	&	2089	&	228, 265					&	10, 20						\\
00031403044	&	1	&	2015-06-21 14:38:59	&	999	&	289								&	20							\\
00031403045	&	1	&	2015-06-22 02:05:58	&	945	&	348								&	10							\\
00031403049	&	5	&	2015-06-22 08:28:58	&	2983	&	379, 381, 448, 506, 514		&	10, 20, 30/10$^*$, 20, 10			\\
00031403047	&	2	&	2015-06-22 09:57:59	&	1909	&	391, 484					&	20, 20						\\
00033832001	&	6	&	2015-06-23 08:00:59	&	1888	&	521, NA, 527, 537, 551, 555&	20, 10, 20, 20, 20, 20		\\
00031403052	&	2	&	2015-06-23 21:23:39	&	279		&	598, 599					&	10, 10	  			        \\
00031403054	&	4	&	2015-06-24 04:58:59	&	5663	&	603, 633, 648, 686			&	20, 10, 20, 20				\\
00031403053	&	1	&	2015-06-24 08:07:58	&	1065	&	640							&	10	   					    \\
00033832002	&	1	&	2015-06-24 11:13:59	&	1530	&	644							&	10       			 		\\
00031403055	&	1	&	2015-06-25 00:03:59	&	1030	&	754 						&	20	   			 			\\
00031403056	&	1	&	2015-06-25 00:21:59	&	820		&	773 						&	20       			 		\\
00031403057	&	3	&	2015-06-25 03:37:59	&	2779	&	798, 826, 858			    &   20, 20, 20         			\\
00031403058	&	1	&	2015-06-25 22:14:59 &	1314	&	893					    	&   20         					\\
00033861001	&	1	&	2015-06-26 12:35:23	&	1500	&	941						&	30							\\
00031403060	&	1	&	2015-06-26 23:46:59	&	1445	&	969						&	20							\\
  \hline 
\end{tabular}\label{tab:log}
\end{table*}																	

\subsection{Data reduction}

V404 Cyg has shown extremely variable emission, with very intense flux peaks spaced by very low flux phases. Additionally, the emission of the X-ray dust scattering halo, seen in the majority of the \textit{Swift} observations of V404 Cyg considered here (see \citealt{Vasilopoulos2016}, \citealt{Heinz2016} and \citealt{Beardmore2016}), might have been contaminating the source spectra.
For these reasons, a careful data reduction must be performed on the \textit{Swift}/XRT data from V404 Cyg in order to obtain consistent results.  We report the data reduction details in Appendix \ref{App:analysis}, while we summarize here the procedure that we adopted.

\begin{itemize}

\item Following the XRT reduction threads,\footnote{http://www.swift.ac.uk/analysis/xrt/\#abs} we first extracted only grade 0 events, which helps minimise the effects of pile-up when the source is bright (i.e. count rate larger then $\sim 150\,\cts$), and reduces the spectral distortion encountered in WT mode below 1.0 keV when the spectra are highly absorbed. 

\item We divided each event file thus obtained into time segments of either 16~s duration or long enough to contain about 1600 counts, after mitigating pile-up in each segment by extracting the source spectrum in an annular region centred at the source position with fixed outer radius (10, 20 or 30 pixels\footnote{One pixel in \textit{Swift}/XRT corresponds to 2.36 arcsec.}) and variable, inner radius (see Appendix \ref{App:analysis}, Sec. \ref{sec:datared} and \ref{sec:halo}). 

\item We thus obtained a total of 1054 pile-up corrected and background-subtracted spectra with exposure times ranging between 16 and 224\,s, to which we applied a background spectrum extracted from a routine WT mode calibration observation of RXJ1856.4--3754 performed on March 2015 (exposure time of 17.8 ks, see \citealt{Beardmore2016} and Appendix \ref{App:analysis}, Sec.  \ref{sec:halo}).

\item We grouped each spectrum in order to obtain a minimum number of 20 counts per bin and added a 3 per cent systematic errors, following the prescriptions given in the  \textit{Swift}/XRT data analysis threads. We have fitted independently all the spectra in the 0.6--10.0\, keV energy range. We used the $\chi^2$ statistics in the search for the best fitting model and for the parameters error determination, and quoted statistical errors at the 1$\sigma$ confidence level.

\end{itemize}

\subsection{Data Analysis}\label{sec:analysis}

We fitted the \swift/XRT spectra
using \textsc{xspec} 12.9.0i \citep{Arnaud1996}, applying different models to our data.
Most of the data could be reasonably well described by a single (often highly) absorbed power law and a narrow Gaussian line around the $\sim$6.4 keV Fe-K$_{\alpha}$  line energy (\textsc{gauss}).
However, using an absorbed power law model we obtained values of the photon index $\Gamma$  significantly lower than 1 in about 45 per cent of the spectral fits, not consistent with the spectral slope of the hard X-ray emission observed by \inte, where $\Gamma >1.4$ \citep{Sanchez-Fernandez2017}.
Furthermore, the hydrogen column density often drifted to values significantly lower than the estimated interstellar value in the direction of V404 Cyg, $N_\textrm{H,gal} = 8.3 \pm 0.8 \times 10^{21}\,\textrm{cm}^{-2}$, obtained from the optical reddening of $A_\textrm{V} = 4.0$ \citep{Hynes2009} and assuming $A_\textrm{V} /N_\textrm{H}(10^{21}\,\textrm{cm}^{-2}) = 0.48$ \citep{Valencic2015}.
This indicated that a more complex spectral model was required. 

Since the presence of additional neutral, local absorption was a characteristic feature of the 1989  outburst of V404~Cyg (see, e.g., \citealt{Oosterbroek1996} and \citealt{Zycki1999}), we added to our model a neutral absorber partially covering the source (\textsc{TBpcf}).
The two main parameters of \textsc{TBpcf}, both left free to vary in our fits, are the equivalent hydrogen column density $N_\textrm{H}$ and a partial covering fraction $\textrm{PCF}$, which takes into account the non-homogeneity of the absorber. $N_\textrm{H}$ was limited between 0 and 10$^3 \times 10^{22}\,\textrm{cm}^{-2}$, $\textrm{PCF}$ could take any value between 0 and 1. We allowed the power law normalization $N_{\Gamma}$ to vary freely, while we set a soft and hard limit to $\Gamma$ at 0 and 4, respectively.
In addition, we used the \textsc{TBfeo} model for the interstellar absorption, assuming the abundances of \cite{Wilms2000} and the cross-sections of \cite{Verner1996}. We fixed $N_\textrm{H,gal} = 8.3 \times 10^{21}\,\textrm{cm}^{-2}$ for this component.

Residuals around 6.4 keV indicated that a narrow line-like feature was required in about 10\% of our spectra. 
Since such a line is likely produced via reflection by  material distant from the BH (see \citealt{Oosterbroek1996}, \citealt{Oosterbroek1997}, \citealt{King2015}, \citealt{Motta2017}), we used a simple Gaussian to describe it. The line centroid energy was allowed to vary between 5.8 and 7.3 keV, while the line width was fixed to 0.1~keV, since the lines seen by \textit{Chandra}  are narrow (see \citealt{King2015}) and are thus unresolved at the \textit{Swift}/XRT CCD spectral resolution. The line normalization was  left free to vary. 
The model that we will discuss has the form: 

\textsc{TBfeo $\times$TBpcf$\times$(powerlaw + gauss)}

\noindent and provides reasonably good and statistically acceptable fits for our entire sample.  
This model depends on 6 parameters: the column density $N_\textrm{H}$ and partial covering fraction $\textrm{PCF}$ (from the \textsc{TBpcf} component), the power law slope $\Gamma$ and normalization $N_{\Gamma}$ (from the \textsc{powerlaw} component), the line centroid energy $E_\textrm{Line}$ and normalization $N_\textrm{Line}$ (from the \textsc{gauss} component).

We note that a simple absorbed disk-blackbody provided good fits only for a few spectra, and such fits are always statistically disfavoured with respect to the fit performed with a partially absorbed power law model.
We also tested various additional continuum components (\textsc{nthcomp}, \textsc{compps}, both describing the continuum produced by thermal Compton up-scattering of soft X-ray photons), different absorber models (\textsc{zxipcf}, \textsc{swind1}, aimed at modelling a partially covering photo-ionized absorber, without and with a velocity shear, respectively). We also tried to convolve the continuum with reflection models (e.g. with \textsc{reflect}, or with the reflection model within \textsc{compps}).  
The use of different continuum models, however, hardly made any difference and returned very similar values of $\chi^2_{\nu}$, which indicates that the underlying spectrum is a power law\footnote{Both \textsc{nthcomp} and \textsc{compps} deviate only slightly from a pure power-law model in the 0.6--10~keV range, and the quality of our data do not allow such small deviations to be detected.} in the 0.6--10~keV range; i.e. we do not see evidence for a low-energy excess originating from a disc-like spectral component, nor a high-energy cut-off due to very low electron temperatures of the Comptonizing medium (that hence would be detected in the \textit{Swift}/XRT energy band).  

Since V404 Cyg has shown strong reflection features (\citealt{Motta2017} and \citealt{Sanchez-Fernandez2017}, \citealt{Walton2017}), as well as times where absorption and/or reflection were not statistically required by the data \citep{Walton2017}, we performed different fit runs, modelling our spectra with slightly different versions of the same model, i.e. with and without partial covering, with and without an additional absorber, and with and without a reflection component. In the latter case, the model used has the form \textsc{TBfeo$\times$TBpcf$\times$(reflect$\ast$power + gauss)}, and depends on one extra parameter -- the reflection fraction $\textrm{Refl}$ -- with respect to the model described above.

Then, we attempted to establish, for each spectrum, which was the best fit using the Bayesian Information Criterion (BIC, \citealt{Schwarz1978}, \citealt{Liddle2007}).
The BIC can be used to choose among competing models with different numbers of parameters, and it is defined as BIC = $\chi^2 + k \log N$, where $k$ is the number of parameters in the model, and $N$ the number of data points fitted.  We considered a $\Delta\mathrm{BIC} > 10$ from two different fits to the same data as strong evidence for the model with lower BIC being the best fit to the data \citep{Liddle2007}. It is worth noting that the BIC also includes a correction term that involves the number of parameters to prevent over-fitting with complex models, therefore a smaller BIC may indicate a better model fit, a lower number of parameters, or both.
For each fit, we also computed the corresponding $p$-value, setting $p=0.05$ as an additional limit for a satisfactory fit.

The BIC test proved useful when we were comparing fits where we used the partial covering absorber and fits where we did not use it. The results of such fits will be described in Sec. \ref{sec:evolution}. However, we note that in the \textit{Swift} energy  range it is almost always very difficult to constrain the reflection component (when present), therefore the use of a reflection component caused fitting problems in a large number of spectral fits (i.e. the reflection fraction and/or the $\Gamma$ were not constrained). This issue is displayed and described in Fig.~\ref{fig:modello}. 

Based on the above, we concluded that phenomenological fits (absorbed power law fits, with or without additional partial covering as determined through the BIC) \emph{not} including the reflection component were the ones least affected by modelling degeneracies, and therefore the most suited to describe the source emission from a phenomenological point of view. Hence, our discussion will be based on the results from such fits.
For the sake of comparison and completeness, we report in Appendix \ref{App:AppendixA} the results of the fit obtained by convolving the partially-covered absorbed power law model described above with a reflection component (\textsc{reflect} model), which in a few cases seems to be the best description of the data. 

\section{Results}\label{Sec:results}

In order to compare the \swift/XRT data with the overall V404 Cyg outburst, we extracted the \textit{INTEGRAL}/IBIS-ISGRI light curve, which provides the best (hard) X-ray coverage of the source for the 2015 June outburst (see \citealt{Kuulkers2015c}). 
Fig.~\ref{fig:ISGRI} (top panel) shows the  \textit{INTEGRAL}/IBIS-ISGRI light curve in the 20--200 keV band (top panel). The Figure also reports the \textit{Arcminute Microkelvin Imager (AMI)} radio light curve (grey points) with a 0.02 days time resolution at 14.6 GHz (Fender et al. in prep).
The bottom panel of Fig.~\ref{fig:ISGRI} shows the \swift/XRT light curve in the 2--10~keV band. The \textit{INTEGRAL}/IBIS-ISGRI and the \swift/XRT light curves  are colour coded according to flux: the redder the points, the higher the flux. In the \textit{INTEGRAL}/IBIS-ISGRI light curve, dark red corresponds to $\sim$60~Crab and light green points corresponds to about 50~mCrab. 
In the \textit{Swift} light curve, dark red points correspond to $\sim$50~Crab, while dark blue points correspond to 50~mCrab. Note that the colour scale is different for the two panels as \textit{Swift}/XRT and \textit{INTEGRAL}/IBIS-ISGRI cover different energy ranges. In the case of IBIS-ISGRI the conversion into Crab units has been done by directly comparing the Crab count rate with that from V404 Cyg during its outburst, while we converted the measured flux directly into Crab units (1 Crab =  $2.4\times 10^{−8} \ergcms$ in the energy range from 2 to 10 keV) in the case of \textit{Swift}/XRT. 
For reference, we indicate in the bottom panel of Fig.~\ref{fig:ISGRI} the spectrum number for the first spectrum extracted from each \textit{Swift} snapshot (unless two adjacent snapshots are too close). 

In both panels, the red arrows mark the eight ejection times reported by \cite{Tetarenko2017}, based on Very Large Array, the Sub-millimeter Array and the James Clerk Maxwell Telescope data taken on MJD 57195 during four hours of observations. Two of these ejections occurred during our \swift/\textit{XRT} data, in particular in correspondence to spectra 470 and 484.
\begin{figure*}
\centering
\begin{tabular}{c c}
Changes in partial covering fraction (144s) &   Changes in column density (80s)\\
\includegraphics[width=0.45\textwidth]{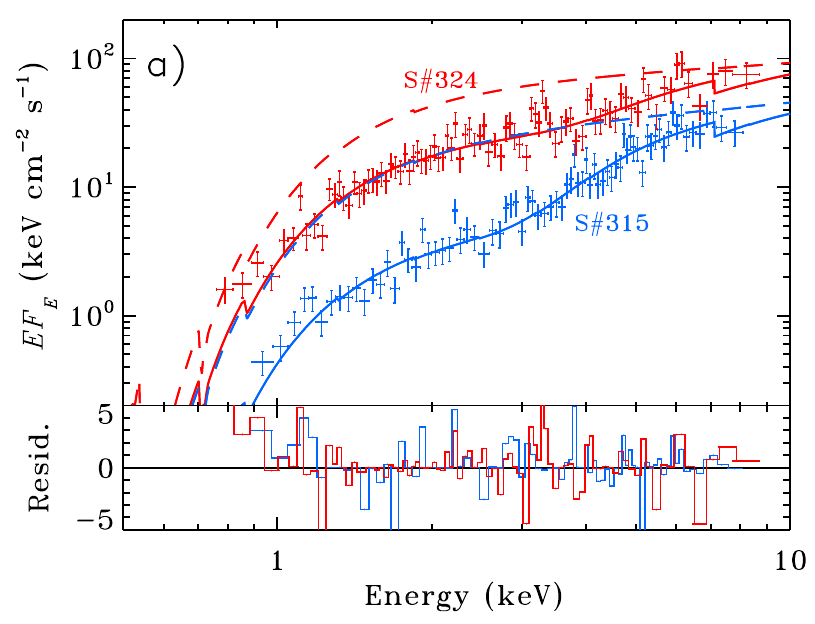}  & \includegraphics[width=0.45\textwidth]{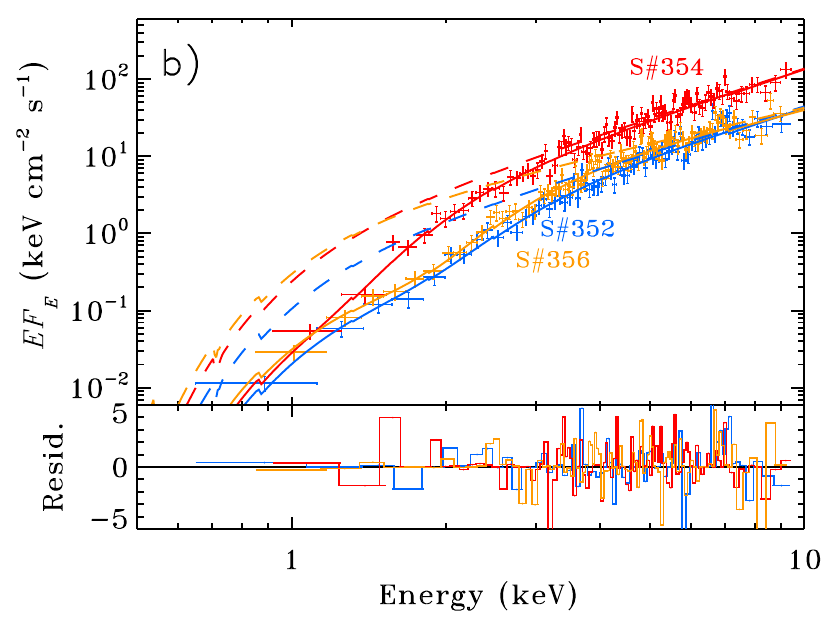} \\
Changes in photon index and partial covering (48s) & Changes in photon index,  partial covering and column density (336s)\\
\includegraphics[width=0.45\textwidth]{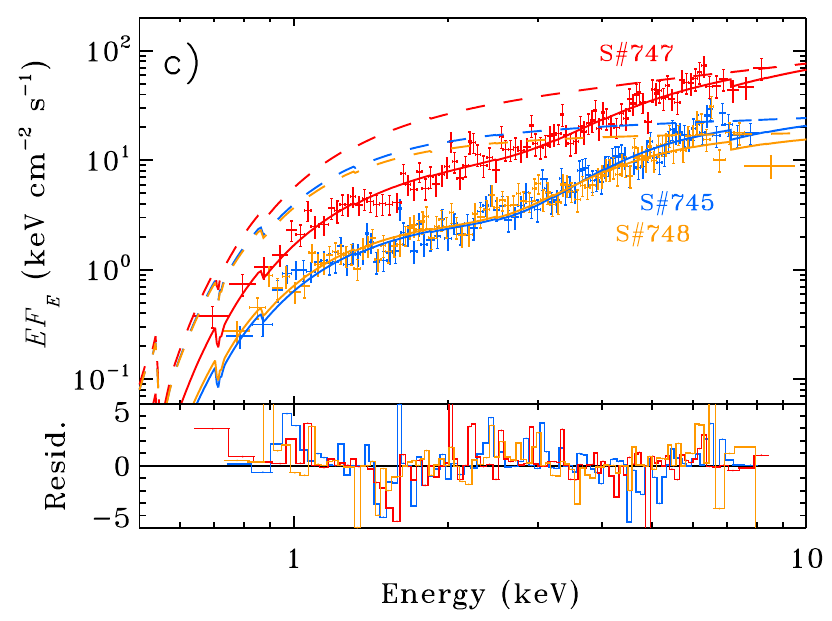}  & \includegraphics[width=0.45\textwidth]{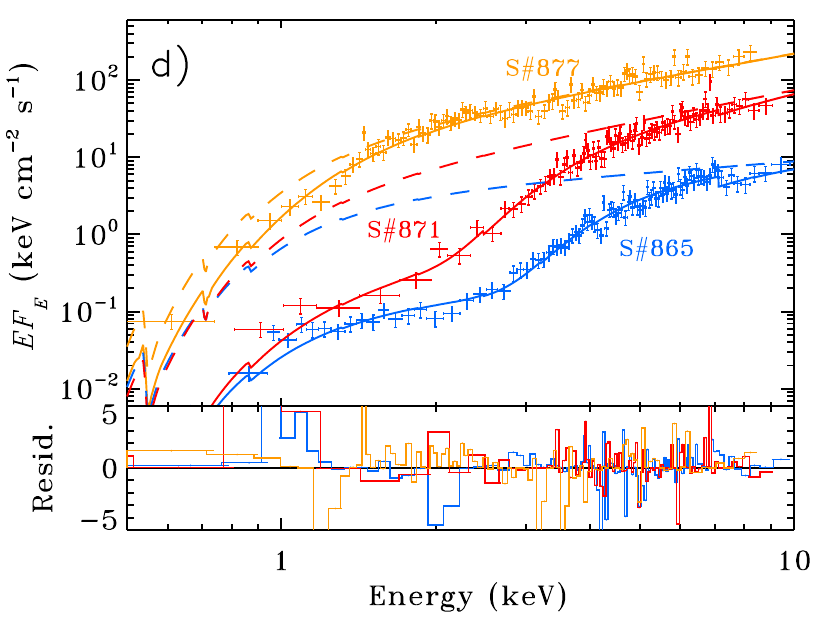} \\
Changes in photon index (448s) &  Examples of spectra showing a narrow line \\
\includegraphics[width=0.45\textwidth]{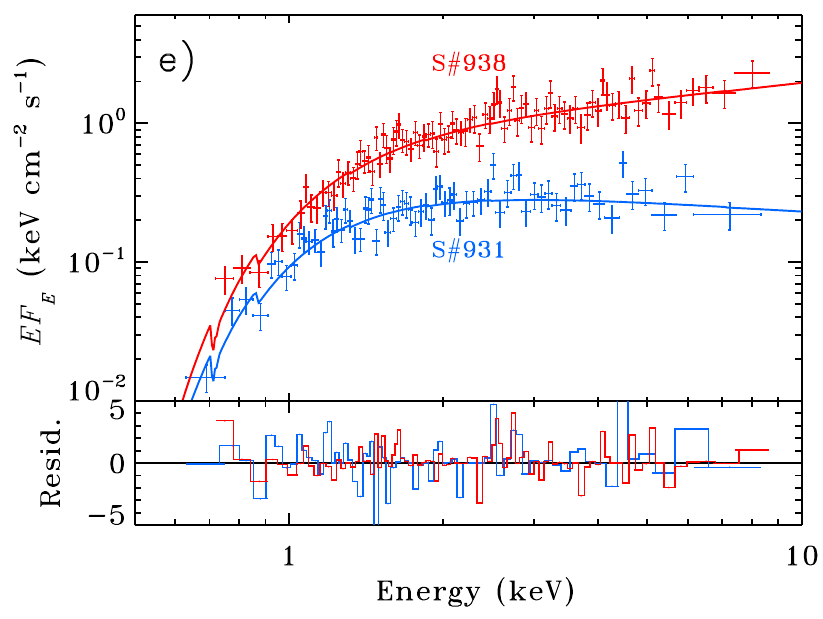} & \includegraphics[width=0.45\textwidth]{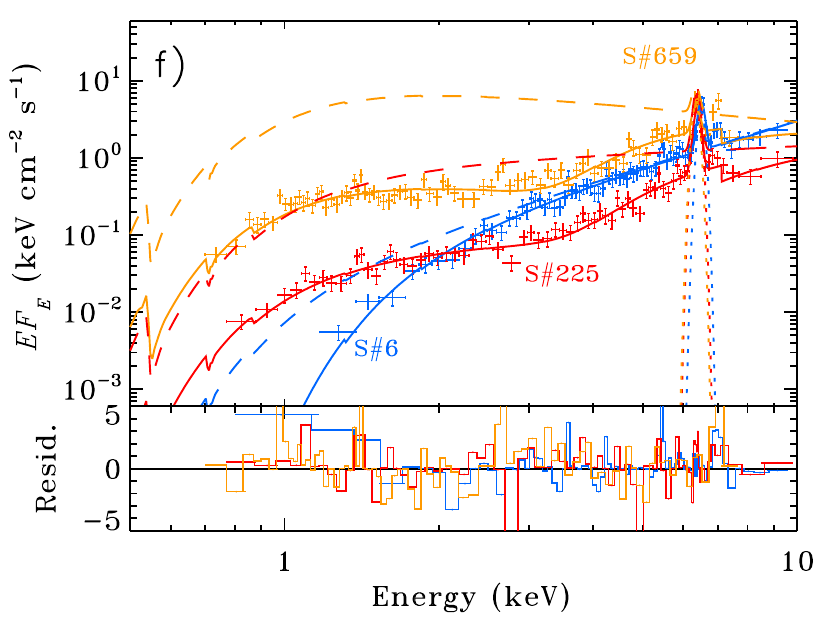} \\
\end{tabular}
\caption{Gallery of rapid spectral variability cases. In each panel we report examples of different kinds of spectral variability observed during the outburst. We plot in the top halves of each panel the spectra, their best fit (solid line) and the unabsorbed model (dashed lines) - where we essentially removed the effect of the local (partial) absorption -  and in the bottom halves the residuals to the best fit. We also give the spectrum number corresponding to each spectrum displayed. Spectra are colour coded as a function of time, according to the sequence blue, red, orange. A brief description of the kind of variability displayed is given on top of each panel, together with the time span covered by the spectra considered (see the text for a detailed description). The best fit parameters for all the spectra shown are reported in Tab. \ref{tab:spectral_fits}. A colour version of this Figure is available on-line.
}
\label{fig:spectra}
\end{figure*}

\subsection{Rapid spectral variability}\label{Sec:spectralvariability}

The \textit{Swift}/XRT spectra that we extracted show remarkable spectral changes on short time-scales. 
Figure \ref{fig:spectra} shows a collection of unfolded spectra that illustrate all the different kinds of rapid spectral variability that we observed. Significant differences between the spectra in each panel are also observed in the folded spectra. Any other case of fast spectral change can be associated with one of the rapid variability cases described below. 
In each panel, the spectra are colour-coded as a function of time: blue comes first, then red and then orange (when present). In each panel, we show the data and their best fit (solid line), and the model corrected for absorption (dashed line). In Tab. \ref{tab:spectral_fits} we report the parameters of the best fits to the spectra displayed. 

\begin{table*}									
\centering									
\caption{V404 Cyg spectral parameters for the fits displayed in Fig. \ref{fig:spectra}: local column density ($N_\textrm{H}$), partial covering fraction ($\textrm{PCF}$), photon index ($\Gamma$), power law normalization ($N_{\Gamma}$), the line energy ($E_\textrm{Line}$) and normalization ($N_\textrm{Line}$), $\chi^2_\nu$ and degrees of freedom (D.O.F.) of the spectra. }  
\begin{tabular}{p{1.5cm} p{2.5cm} p{2.5cm} p{2.5cm} p{2.5cm}}														
\hline \hline			
Panel	&	parameter	&	blue sp.	&	red sp.	&  orange sp.\\
\hline		\hline				
{\bf a)} &	$N_\textrm{H}$ ($10^{22}$ cm$^{-2}$)	&	21$\pm$3	&	32$_{-10}^{+15}$	&	 	\\
		&	$\textrm{PCF}$		    &	0.82$\pm$0.04	&	0.59$_{-0.09}^{+0.08}$		&	 	\\
		&	$\Gamma$	&	1.6$\pm$0.2 	&	1.7$\pm$0.1		&	 	\\
        &	$N_{\Gamma}$ &	20$_{-5}^{+7}$	&	52$_{-11}^{+13}$		&	    \\
        & 	$\chi^2_\nu$	 &	81.63			&	104.98					&		\\
        &	D.O.F.		 & 	73				&	83						&		\\
\hline				
{\bf b)}	&	$N_\textrm{H}$ ($10^{22}$ cm$^{-2}$)		&	9$\pm$2	&	4$\pm$1	    &	10$\pm$2	\\
		&	$\textrm{PCF}$			&	0.81$\pm$0.03		&	0.88$_{-0.04}^{+0.05}$	&	0.89$_{-0.03}^{+0.02}$	\\
		&	$\Gamma$	&	0.0 (fixed)	&	0.0 (fixed)	&	0.4$\pm$0.2	\\
        &	$N_{\Gamma}$ &	0.44$\pm$0.02		&	1.28$_{-0.04}^{+0.05}$		&	1.3$_{-0.3}^{+0.4}$		\\
        & 	$\chi^2_\nu$	 &	79.16		&	104.76		&	70.42				\\
        &	D.O.F.		 & 	74			&	99			&	73					\\
\hline				
{\bf c)}		&	$N_\textrm{H}$ ($10^{22}$ cm$^{-2}$)		&	28$_{-3}^{+4}$	&	27$_{-5}^{+8}$	&	24$_{-3}^{+4}$		\\
		&	$\textrm{PCF}$			&	0.84$\pm$0.03	&	0.70$_{-0.06}^{+0.05}$	&	0.80$\pm$0.04	\\
		&	$\Gamma$	&	1.8$\pm$0.1 	&	1.5$\pm$0.1 	&	1.9$\pm$0.1				\\
        &	$N_{\Gamma}$ &	17$_{-3}^{+5}$	&	24$_{-5}^{+7}$	&	16$\pm$4    			\\
        & 	$\chi^2_\nu$	 &	95.07		&	83.57		&	97.22							\\
        &	D.O.F.		 & 	90			&	90			&	90								\\
\hline				
{\bf d)}	&	$N_\textrm{H}$ ($10^{22}$ cm$^{-2}$)		&	34$\pm$2	&	17$\pm$2	&	0.9$\pm$0.5					\\
		&	$\textrm{PCF}$			&	0.96$\pm$0.01	&	0.95$\pm$0.01	&	0.6$_{-0.2}^{+0.3}$ 	\\
		&	$\Gamma$	&	1.5$\pm$0.1		&	0.7$\pm$0.2	&	0.8$\pm$0.1				\\
        &	$N_{\Gamma}$ &	3$\pm$1			&	4$_{-1}^{+2}$		&	14$_{-2}^{+3}$ 		\\
        & 	$\chi^2_\nu$	 &	108.48			&	 85.45				&	113.64				\\
        &	D.O.F.		 & 	100				&	 87					&	95					\\
\hline				
{\bf e)}	&	$N_\textrm{H}$ ($10^{22}$ cm$^{-2}$)		&	-	&	-	&	 	\\
		&	$\textrm{PCF}$			&	-	&	-	&	 	\\
		&	$\Gamma$	&	2.22$\pm$0.05	&	1.61$\pm$0.04	&	 	\\
        &	$N_{\Gamma}$ &	0.39$\pm$0.02	&	0.80$\pm$0.03	&	   \\
        & 	$\chi^2_\nu$	 &	78.20			&	71.37 			&		\\
        &	D.O.F.		 & 	68				&	90				&		\\
\hline				
{\bf f)}	&	$N_\textrm{H}$ ($10^{22}$ cm$^{-2}$)	&	2.0$\pm$0.3	&	61$_{-9}^{+10}$	    				&	53$_{-8}^{+9}$	     			\\
		&	$\textrm{PCF}$				&	-									&	0.92$\pm$0.02						&	0.93$\pm$0.01					\\
		&	$\Gamma$		&	0 (fixed)							&	1.7$_{-0.1}^{+0.1}$ 				&	2.62$\pm$0.16					\\
        &	$E_\textrm{Line}$  &	6.50$\pm$0.04						&	6.38$\pm$0.02						&	6.34$_{-0.09}^{+0.08}$			\\
        &	$N_\textrm{Line}$  &	(2.4$\pm$0.3)$\times$10$^{-2}$	&	(4.2$\pm$0.5)$\times$10$^{-2}$    		&	(3$\pm$1)$\times$10$^{-2}$ \\
        & 	$\chi^2_\nu$	 	&	67.31								&	84.98 								&	110.66							\\
        &	D.O.F.		 	& 	76									&	71									&	77								\\
\hline				

\end{tabular}\label{tab:spectral_fits}									
\end{table*}																							

\begin{itemize}
\item Panel a) shows a fast change in the partial covering fraction. The two spectra cover a 144\,s span. The blue spectrum is characterized by a high covering fraction (82\%) and moderate local column density ($N_\textrm{H} \approx 2\times10^{23}\,\cm2$), while the red spectrum, shows a lower covering fraction (60\%) and a slightly higher local column density ($N_\textrm{H} \approx 3\times10^{23}\,\cm2$). The photon index is consistent with being constant through the spectral change, remaining in the $\Gamma \approx 1.6$--$1.7$ range. We note that the column density is not well constrained for the red spectrum, but it is consistent with that of the blue spectrum. This is probably a consequence of the partial covering factor being lower than in the blue spectrum ($N_\textrm{H}$ and \textrm{PCF} can be degenerate parameters in the case where one or the other parameter is close to or consistent with zero). This spectral change might correspond to a situation where the source intrinsic emission is first partially obscured by dense matter -- possibly a clump -- and then unveiled as the matter moves away from the line of sight, causing a sudden drop in covering fraction, and revealing the emission from the inner accretion flow.

\item Panel b) shows a case where major changes in the spectral shape are ascribed to fast changes in the local column density, while the partial covering fraction and the intrinsic spectrum remain consistently the same. In the three spectra the covering fraction is consistent with 80--90\%, but while in the blue and orange spectrum the column density is about $N_\textrm{H} \approx 10^{23}\,\cm2$, in the red spectrum the local column density is about a factor of 2 lower. As can be seen from the models corrected for absorption, the de-absorbed spectrum varies only very slightly. The change in the column density shown in the figure appears as a flare in the light curve of the source, possibly a consequence of the fact that the material covering the source becomes suddenly less dense. 

\item Panel c) shows a case of a fast hard flare. The three spectra shown in the figure are extracted from adjacent time stretches and cover a time span of only 48~s. While the blue and orange spectra show a photon index of about $\Gamma \approx 1.8$--$1.9$, moderate column density ($N_\textrm{H} \approx [2-3]\times 10^{23}\,\cm2$), and relatively high covering fraction (84\% and 80\%, respectively), the red spectrum -- corresponding to the flare -- shows significantly harder photon index ($\Gamma \approx 1.5$), lower covering fraction (70\%), but column density consistent with the blue and orange spectra ($N_\textrm{H} \approx 3\times10^{23}\,\cm2$). This fast spectral change likely corresponds to a situation where a clump that might be partially obscuring the source moves slightly away from the line of sight, uncovering a particular region of the inner accretion flow that is emitting a hard spectrum (see Sec. \ref{sec:discussion}). 

\item Panel d) shows a situation similar to those described for Panel b and c, though somewhat more extreme. This spectral change occurs in about 6 minutes (336~s) and in this case the blue spectrum is characterized by a photon index of $\Gamma \approx 1.5$, is highly covered (96\%) and affected by moderate local column density ($N_\textrm{H} \approx 3.4 \times 10^{23}\,\cm2$). The red spectrum, instead, is harder (with $\Gamma \approx 0.7$), still highly covered (95\%), but affected by a local column density a factor of two lower than in the case of the blue spectrum. Finally, the orange spectrum -- corresponding to a flare -- is affected by low local column density  ($N_\textrm{H} \approx 10^{22}\,\cm2$) covering 64\% of the emission from the inner accretion flow. This situation could be the result of a decrease in the local column density, i.e. maybe a consequence of the obscuring material getting thinner. This is accompanied by a partial uncovering of a region of the inner accretion flow (with a hard spectrum), which results in a significant increase in flux. 

\item Panel e) shows a clear case of variations of the photon index, while the source is not affected by significant covering and/or local column density. The change displayed occurs on a time scale of a few minutes (448~s) and corresponds to an actual spectral change of the source, that hardens while increasing in flux. While in the blue spectrum the photon index is $\Gamma \approx 2.2$, the red spectrum is significantly harder, with $\Gamma \approx 1.6$. 

\item Panel f) shows a collection of spectra with a line feature in the iron-K$\alpha$ line region, collected at different times during the outburst. The blue spectrum, taken at the beginning of the outburst, shows a prominent but narrow Gaussian line in emission in the iron line region, and low, uniform absorption (i.e. no partial covering is statistically required). Interestingly, this spectrum appears extremely hard, with a photon index consistent with 0. The red spectrum shows the effects of high column density and of the presence of partial covering. Also in this case, a prominent line is evident. The spectral photon index is significantly higher than in the previous case ($\Gamma \approx 1.7$). The orange spectrum is much softer than the previous ones, requiring a photon index of $\Gamma \approx 2.6$. Here the Gaussian line is slightly weaker than in the red spectrum described above, though still clearly present.  Also in this spectrum the effects of the relatively high absorption and covering are evident. 

\end{itemize}
\begin{figure*}
\centering
Parameter evolution from partially-covered, absorbed power law.
\includegraphics[width=0.94\textwidth]{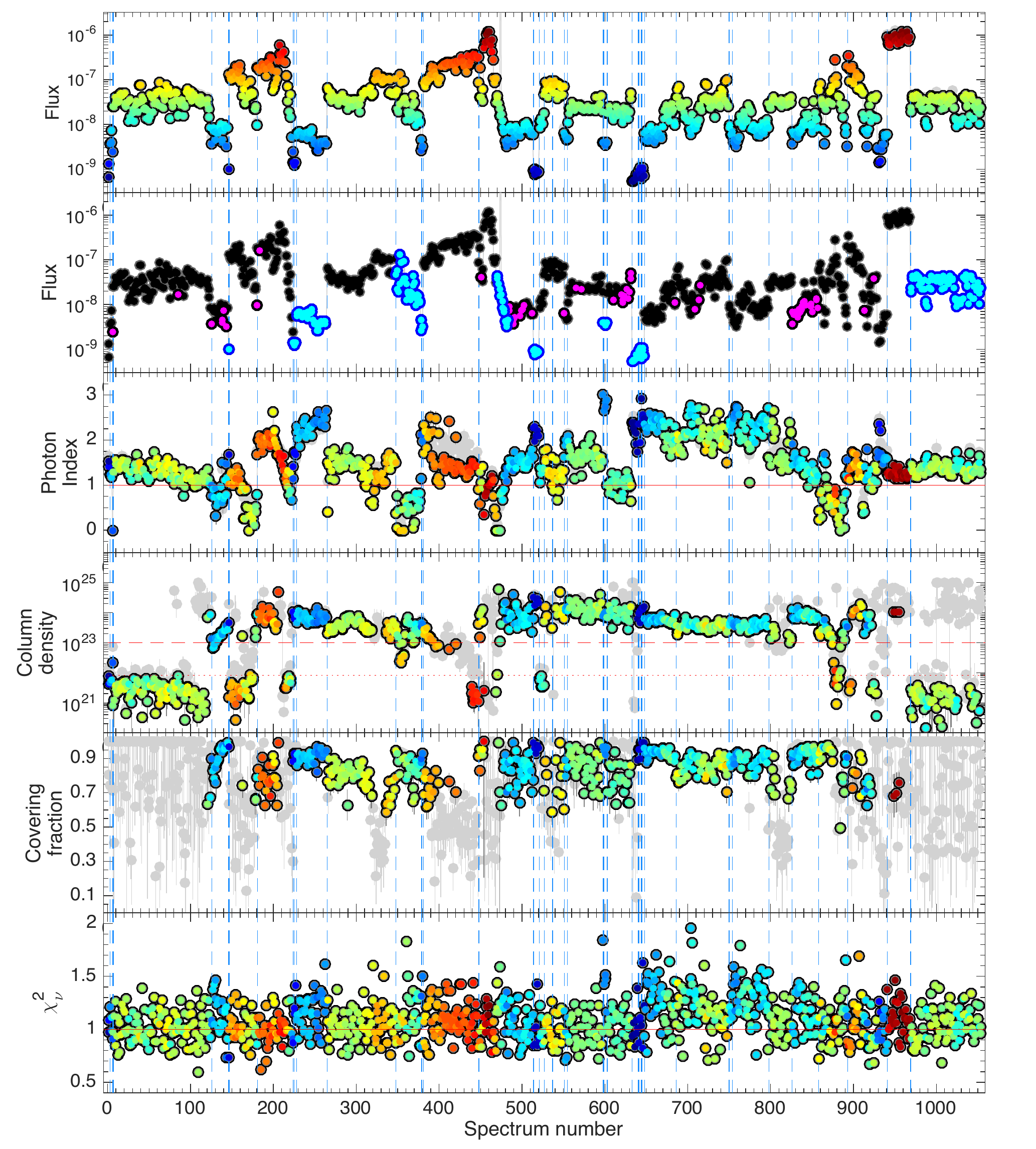}
\caption{Time evolution of the best fitting parameters obtained modelling the spectra of our sample with a partially-covered, absorbed power law, with the addition of a Gaussian line in emission. All parameters are plotted as a function of spectrum number instead of time to facilitate parameter inspection. The beginning/end of each \textit{Swift} snapshot is marked by a blue dashed line. In all panels save for the second from the top, points are colour-coded as a function of flux (in $\ergcms$), shown in the top panel. Other panels show: flux (in units of $\ergcms$) corrected for the ISM absorption, with magenta indicating spectra that show a narrow Gaussian line in emission, and cyan indicating spectra where the emission from the dust scattering halo might be affecting the source spectrum; spectral photon index; intrinsic column density; partial covering fraction; $\chi^2_\nu$. The red line marks photon index equal 1 in the photon index panel, column density equal 10$^{23}$cm$^{-2}$ (dashed) and column density equal to the ISM value (dotted) in the column density panel,  and $\chi^2_\nu$ equal 1 in the $\chi^2_\nu$ panels. 
Note that in the partial covering fraction panel, for clarity we only plotted the points indicating covering fractions less than 1, as covering fraction equal 1 effectively means that the absorber is uniform.  
The grey points show the parameter evolution obtained leaving the covering fraction and the local column density always free to vary, while the coloured points show the results of the fits to the spectra of our sample after applying the BIC test in order to determine whether the partial covering was statistically required (see Sec. \ref{sec:analysis} for details).
A colour version of this Figure is available on-line.}
\label{fig:parameters}
\end{figure*}


\subsection{Parameter evolution}\label{sec:evolution}

In Fig. \ref{fig:parameters} we show the results of the spectral fits described in Section \ref{sec:analysis}. Since the time gaps between different \textit{Swift} snapshots are on average much larger than the typical length of a snapshot, we show the variations of the spectral parameters as a function of spectrum number, instead of as a function of time, in order to make the parameter inspection easier.  
The beginning/end of each \textit{Swift} snapshot is marked by a blue dashed line in  Fig. \ref{fig:parameters}. 
The \textit{Swift} observations are considerably sparser than the \inte\ ones (see Fig. \ref{fig:ISGRI}), however the length of \textit{Swift} and \inte\ snapshots was comparable (1~ks and 2-3~ks, respectively for \textit{Swift} and \inte).Both the \textit{Swift} and \inte\ light-curves show large variations in flux, up to three orders of magnitude occurring on time scales from less than a minute up to days.

In Fig.  \ref{fig:parameters}  we show, from top to bottom: observed flux in the 2--10~keV band (in units of $\ergcms$); observed flux in the 2--10~keV band, highlighting in magenta colour those spectra which showed an iron line around 6.4 keV and in cyan the spectra extracted in time intervals where the dust scattering halo is seen around the source in the \textit{Swift}/XRT field of view; photon index ($\Gamma$); local column density ($\NH$); partial covering fraction ($\textrm{PCF}$) and the $\chi^2_\nu$.  
By observed flux we refer to the flux measured in the 2--10 keV energy band, corrected only for interstellar absorption.
All the points in each panel (excluding the second panel from the top) are colour-coded according to flux, where red corresponds to the highest fluxes and blue correspond to the lowest fluxes. 
The grey points are shown for reference and mark the parameter evolution obtained leaving the covering fraction and the local column density always free to vary, while the coloured points show the results of the fits to the spectra of our sample after applying the BIC test in order to determine whether the partial covering was statistically required (see Sec. \ref{sec:analysis}).

The results of the spectral analysis were also used to produce Fig. \ref{fig:corner}, which shows the parameter and observed flux distributions and correlations. From top to bottom (left column) and from left to right (bottom line) we show: local column density $\NH$ (in log scale), partial covering fraction, photon index $\Gamma$, apparent flux (in log scale) and line energy. At the top of each distribution we report the average value of each parameter. 

\subsubsection{Flux}\label{Sec:flux_evolution}

V404 Cyg showed a very large  dynamical range in flux during the June 2015  outburst, with flux variations of almost four orders of magnitude (Fig. \ref{fig:parameters}, top panel).
In particular, in three outburst phases, corresponding to three of the brightest flares caught by \textit{Swift},  the flux changes are particularly remarkable. Around spectra 200--220, 460--490 and 890--910 we can observe a change in flux of 2.4, 2.5 and 2 orders of magnitude in a matter of 290, 400 and 330 seconds, respectively. 
Each of  these three flares is associated with  spectral changes, and in particular to a significant hardening of the source spectrum, which also corresponds to a dramatic drop in column density and to the disappearance of inhomogeneity in the absorber (i.e. the covering factor becomes solidly consistent with one). 

Similar flux variations accompanied by spectral hardening and decrease in $\NH$ occur also at even shorter time-scales, i.e. tens of seconds (see Panel C of Fig. \ref{fig:spectra} and Sec. \ref{Sec:spectralvariability}), where again an increase in flux corresponds to a spectral hardening.

\subsubsection{Photon index}

Over the outburst the photon index varies between $\Gamma \approx 0$ and 3. We note that there are 5 epochs where the photon index $\Gamma$ drifts significantly below 1 (i.e. points below the red solid line in Fig. \ref{fig:parameters}, photon index panel). Despite the very low values reached by this parameter, $\Gamma$ is always well constrained. We remind the reader that $\Gamma$ was
constrained to be larger than zero. However, even in those cases where $\Gamma$ is  consistent with zero (in 1.4\% of our spectra), if left free to vary, it remains well constrained and does not show degeneracy with the $N_\textrm{H}$.
Analysis of the \textit{INTEGRAL}/IBIS-ISGRI data in the 20--200 keV clearly shows that there are no spectra with photon indices below $\Gamma < 1.4$ \citep{Sanchez-Fernandez2017}, and it is therefore likely that for these epochs the partially-covered absorbed power law spectral model is too simplistic. We also note that in $\sim$20\% of the spectra $\Gamma$ is consistent with  or higher than $\sim$1.4, even though fits performed including a reflection component provide a better description to the data. A clear case can be seen comparing the photon index of spectra 650--800 in Fig. \ref{fig:parameters} and in Fig. \ref{fig:parameters_refl}, where we see that the introduction a reflection component reduces the $\chi^2_\nu$. The addition of a reflection component in such spectra causes a drop in the $\chi^2_{\nu}$ from an average value of $\approx$99.40/81 to $\approx$90.05/80\footnote{Note that the number of degrees of freedom is variable from spectrum to spectrum. Therefore, here we are quoting average values for both the $\chi^2$ and the number of degrees of freedom.}, which corresponds to a marginal improvement of the fits. These cases, similarly to those cases where $\Gamma$ is significantly lower than 1, are found at times where the column density and the covering fraction are both high. 

Figure \ref{fig:parameters} also shows that the highest photon indices almost always correspond to the lowest observed fluxes, even though very low values of  $\Gamma$ appear at various phases of the outburst. 
Overall, there is no obvious correlation between the variations in the photon index and the changes in luminosity.  
This behaviour is notably different from that seen in other BHBs, where an increase in flux typically corresponds to constancy or increases in the photon index (see e.g. \citealt{Dunn2010}). However, a similar behaviour has also been seen in the V404 Cyg \inte\ data taken during this same outburst \citep{Sanchez-Fernandez2017}.

\subsubsection{Variable intrinsic absorption}\label{Sec:Absorption}

The local absorption, often characterized by high column density and partial covering fraction, shows very intriguing evolution during the outburst. 
Figure \ref{fig:parameters} (4th panel from the top) shows that the majority of the high column density measurements come from the phase of the outburst where most of the flaring is observed (spectra 130--969), while much lower values ($N_\textrm{H}\sim 10^{21}\,\cm2$), come from the first three observations (spectra 1--129), from the last one (969--1054), and from isolated dips where the column density drops suddenly for  short periods of time (see, e.g., spectra 430--460 in Fig. \ref{fig:parameters}). We stress that the ISM column density is taken into account through a separate model component, thus the low-column density values shown in the plot correspond to low density absorbing material \emph{additional} to that due to the ISM.

Comparing the column density panel and covering fraction panel in Fig. \ref{fig:parameters}, one can see that the absorber shows a bi-modal behaviour: when the column density associated with the local absorber is high, the absorber is highly inhomogeneous, with high and well-constrained values of the covering fraction. Instead, when the column density is low, the absorber is homogeneous (no covering fraction is statistically required by the fits, which from the modelling point of view means covering fraction = 1). Since the column density and partial covering fractions are well constrained, we exclude that this effect could be due to degeneracy in the fits.
The high-column density phases are only occasionally replaced by clear $N_\textrm{H}$ dips, which are often  accompanied by changes in the covering fraction (see also Fig. \ref{tab:spectral_fits}, panel B). For instance, a prominent case where $N_\textrm{H}$ drops from about $2\times 10^{23}\,\cm2$ to $N_\textrm{H} \lesssim 10^{21}\,\cm2$, and the partial covering fraction varies by a factor of 2 (from 0.5 to 1) in a matter of a few minutes can be seen around spectra 320--340.

By comparing the photon index panel with the column density and covering fraction panels, we note that while the photon index has a long-term evolution that does not seem to correlate with the absorber, most of the minute and sub-minute time-scale variations in flux/photon index could be related to changes in local column density and/or covering fractions.

\subsubsection{Variable line}\label{Sec:line}

\begin{figure*}
\centering
\includegraphics[width=1.0\textwidth]{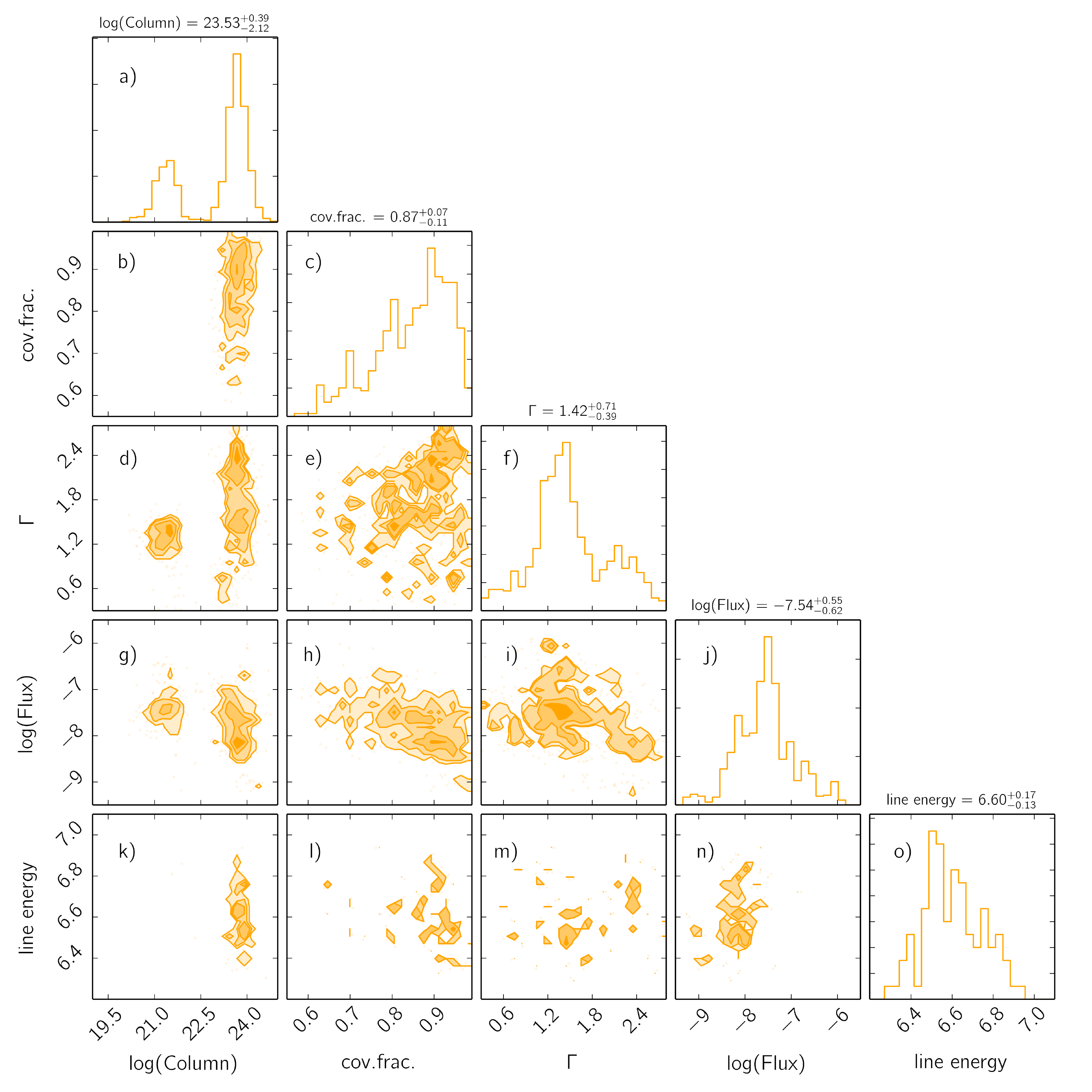}
\caption{Parameter distributions, observed flux distribution, and parameters correlations from the best fits to the spectra of our sample using the partially-covered absorbed power law model described in Sec. \ref{sec:analysis}. From top to bottom (left column) and from left to right (bottom line) we show: intrinsic column density (in log scale), partial covering fraction, photon index $\Gamma$, apparent flux (in log scale, corrected for the ISM absorption) and iron line energy. At the top of each distribution we report the mean value of each parameter.}
\label{fig:corner}
\end{figure*}

A narrow Gaussian lines in emission around 6.4 keV is detected at certain times during the outburst. In particular, it seems to be associated with low to moderate flux levels  (strictly below $2\times10^{-7}\,\ergcms$, see Fig. \ref{fig:parameters}, second panel from the top) and appears during times when the measured column density and covering fraction are high (i.e. above $\sim$10$^{23}$ cm$^{-2}$ and 0.85, respectively). This is particularly evident in the line energy versus $N_\textrm{H}$ panel of Fig. \ref{fig:corner} (panel k), where one can see that no line is detected at column densities lower than $\sim$3 $\times$ 10$^{22}$ cm$^{-2}$.

Even though the photon index associated with the appearance of the line is low to moderate (from below 1 to about 1.5), there is no clear correlation between the presence of the line and the source spectral slope. 

The spectral fits to the data revealed that the lines detected in our spectra are not resolved by \textit{Swift}/XRT. We note that the limited sensitivity of \textit{Swift}/XRT above 8 keV, combined with the presence of edges caused both by heavy absorption and significant reflection, makes it sometimes difficult to properly model the iron lines region, and in particular to constrain the reflection parameters when a reflection component is included in the spectral model. 

\subsection{Spectral parameters distributions}

In Fig. \ref{fig:corner} we show the distributions of the spectral parameters described above, together with the correlations between all the parameters and the observed flux. 

\subsubsection{Flux distribution}

The observed flux distribution (Fig. \ref{fig:corner}, panel j), obtained correcting the 2--10 keV flux only for interstellar absorption, shows a main peak at $\sim$\textrm{3}$\times$10$^{-8}\,\ergcms$. Once converted to bolometric X-ray luminosity,\footnote{We used a bolometric luminosity correction factor equal to 2.2, estimated based on the results reported by \citealt{Sanchez-Fernandez2017}.} this flux corresponds to about 4\% of the expected Eddington luminosity for V404 Cyg (1.6$\times$10$^{-6}\,\ergcms$ for a stellar mass BH of 9 M$\odot$ and at 2.39 kpc, as appropriate for V404 Cyg). However, the flux distribution extends up to 1.22$\times$10$^{-6}\,\ergcms$. The corresponding bolometric luminosity exceeds the Eddington luminosity by a factor of $\approx$2, and a closer look at the flux distribution shows that about 3\% of the spectra return an observed luminosity higher than the Eddington one. It is noticeable that in these cases the spectra did not require any extra column density local to the source, suggesting that the source  might have been completely un-obscured at the time of these super-Eddington events. 

\subsubsection{Photon index distribution}

\begin{figure}
\includegraphics[width=0.45\textwidth]{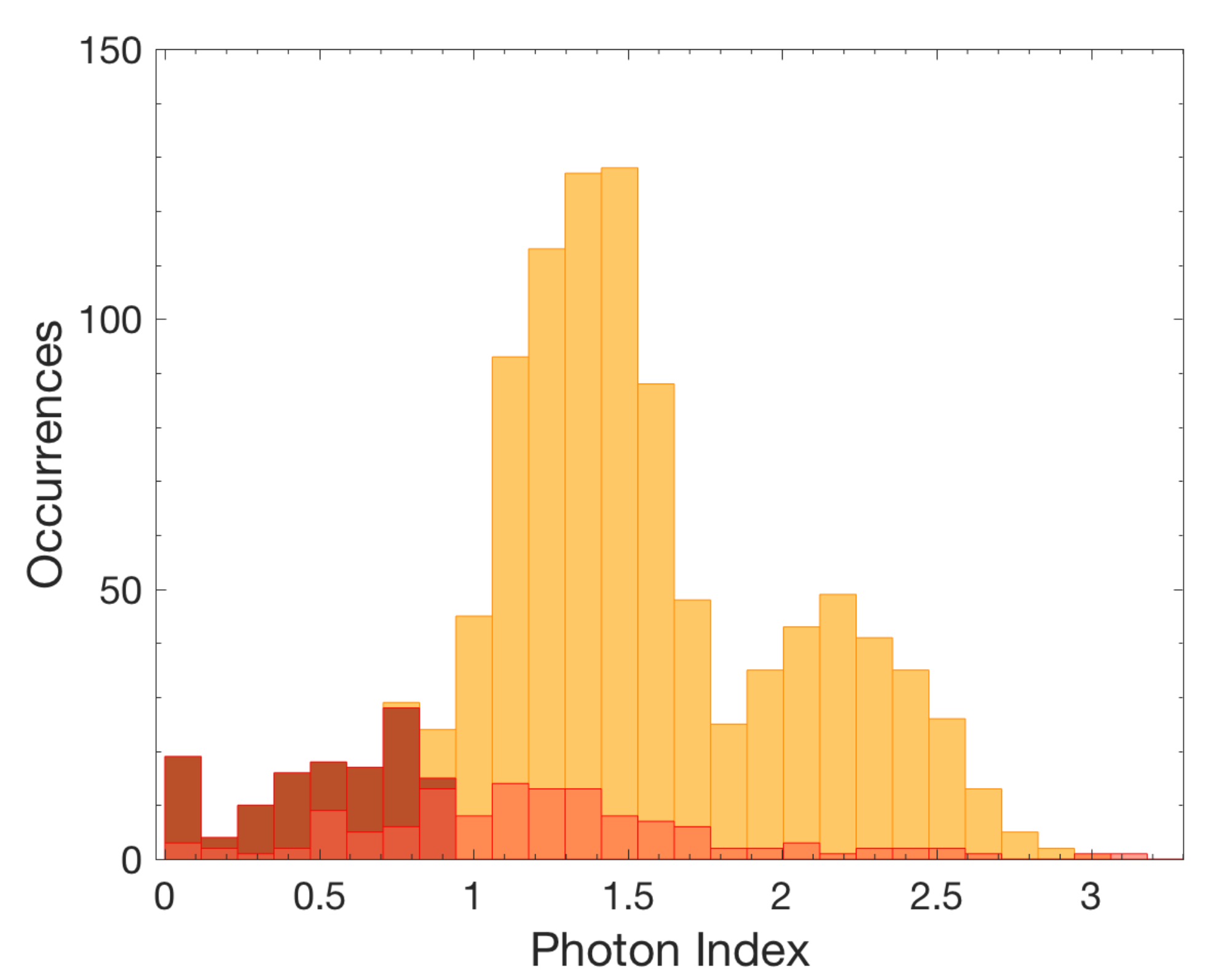}
\caption{Comparison of the photon index distributions obtained when fitting a partially-covered absorbed power law, and a partially covered absorbed power law convolved with a reflection component.  The orange distribution is formed by values of Gamma from a simple absorbed power law fit. The dark red distribution contains all the $\Gamma$ values that are lower than and not consistent with 1. The red distribution is formed by the $\Gamma$ values obtained from the spectra that originally returned $\Gamma$ lower then and not consistent with 1, and that were subsequently modelled with a partially-covered absorbed power law convolved with a reflection kernel. One can see that the addition of the reflection component cause an increase in $\Gamma$ in the majority of cases. A colour version of this Figure is available on-line. }
\label{fig:distributions_mod}
\end{figure}

The $\Gamma$ distribution in Fig. \ref{fig:corner} (panel f) is double-peaked, somewhat similar to what is found with \inte\ \citep{Sanchez-Fernandez2017}. We note, however, that the \textit{Swift} distribution is more spread to the higher and lower values, likely due to the fact that \textit{Swift} lacks sensitivity at high energies, necessary to well constrain the spectral slope. 
By comparing the time evolution of $\Gamma$ obtained using just a partially-covered absorbed power law and the same model convolved with a reflection model, we see that the low-photon index tail is probably due to the presence of reflection in some low-flux spectra that, as noted above, forces the photon index to lower values in order to account for a hard excess. 
This effect is obvious when we substitute the photon index values obtained from the fits of such spectra (about 7\% of the spectra in the sample), with the $\Gamma$ values obtained by convolving our power law model with a reflection component: the photon index moves up to values closer to 1.4. 
This is illustrated in Fig. \ref{fig:distributions_mod}: the orange distribution corresponds to the fits performed with a partially-covered absorbed power law model and is the same $\Gamma$ distribution shown in Fig. \ref{fig:corner} (panel f). The dark red distribution highlights the values of the orange distribution smaller than and not consistent with 1. The light red distribution contains the $\Gamma$ values obtained fitting the spectra that returned $\Gamma$ values lower than and not consistent with 1 with a partially-covered absorbed power law convolved with a reflection model.

\subsubsection{Intrinsic absorption distribution}\label{Sec:Absorption_dist}

The local absorption is highly variable and forms a bi-modal distribution in log scale (see Fig. \ref{fig:corner}, panel a), which reflects the bi-modal evolution that was already evident from Fig. \ref{fig:parameters} (column panel).
Both peaks of the distribution, at $2.5\times10^{21}\,\textrm{cm}^{-2}$ and at $5\times10^{23}\,\textrm{cm}^{-2}$, follow a log-normal distribution. It must be noted that, while the effects of the ISM column density are taken into account separately (fitted by a separate model component, see Sec. \ref{sec:analysis}), the peak at low column density is centred at values of $\sim 10^{21}\,\textrm{cm}^{-2}$. The FWHM of such a peak is $\approx$3.2$\times 10^{21}\,\textrm{cm}^{-2}$, significantly larger than the uncertainty on the ISM column density (equal to $0.8 \times 10^{21}\,\textrm{cm}^{-2}$), and therefore of the expected width of such a distribution if it was due to fluctuations in the best fits. This suggests that the low column density peak of the distribution is a real effect related to an additional, though thin, layer of absorbing material. We stress that the data do not allow to completely exclude that this is an artificial effect of our spectral modelling. 

High column density neutral absorbing local material ($N_\textrm{H} \approx 10^{22} - 10^{25}\, \textrm{cm}^{-2}$), covering a variable fraction of the central X-ray source is necessary to describe 54\% per cent of our spectra. In the remaining 46\% of the spectra, the local absorbing material is either uniform and thin ($N_\textrm{H} \sim 10^{21}\,\textrm{cm}^{-2}$ and $\textrm{PCF} = 1$) or absent.  

Inspecting the column density distribution panel and the covering fraction versus column density panel in Fig. \ref{fig:corner} (panel a and b, respectively), the bi-modal structure of the absorber mentioned in Sec. \ref{sec:evolution} is obvious. On the one hand, only those spectra requiring high column densities (above 10$^{22}\,\textrm{cm}^{-2}$) also require partial covering. On the other hand, none of the spectra that shows low column densities requires a partial covering, again pointing to  the homogeneity of the absorber when the column density is very low. Even though we cannot completely rule out the possibility that this dichotomy arises as a consequence of the fit being insensitive to the partial covering parameter in the presence of low column densities, the fact that the dichotomy is extremely sharp suggests that this effect indeed reflects a real property of the absorber.

\subsubsection{Iron line energy distribution}\label{Sec:line_dist}

\begin{figure}
\centering
\includegraphics[width=0.45\textwidth]{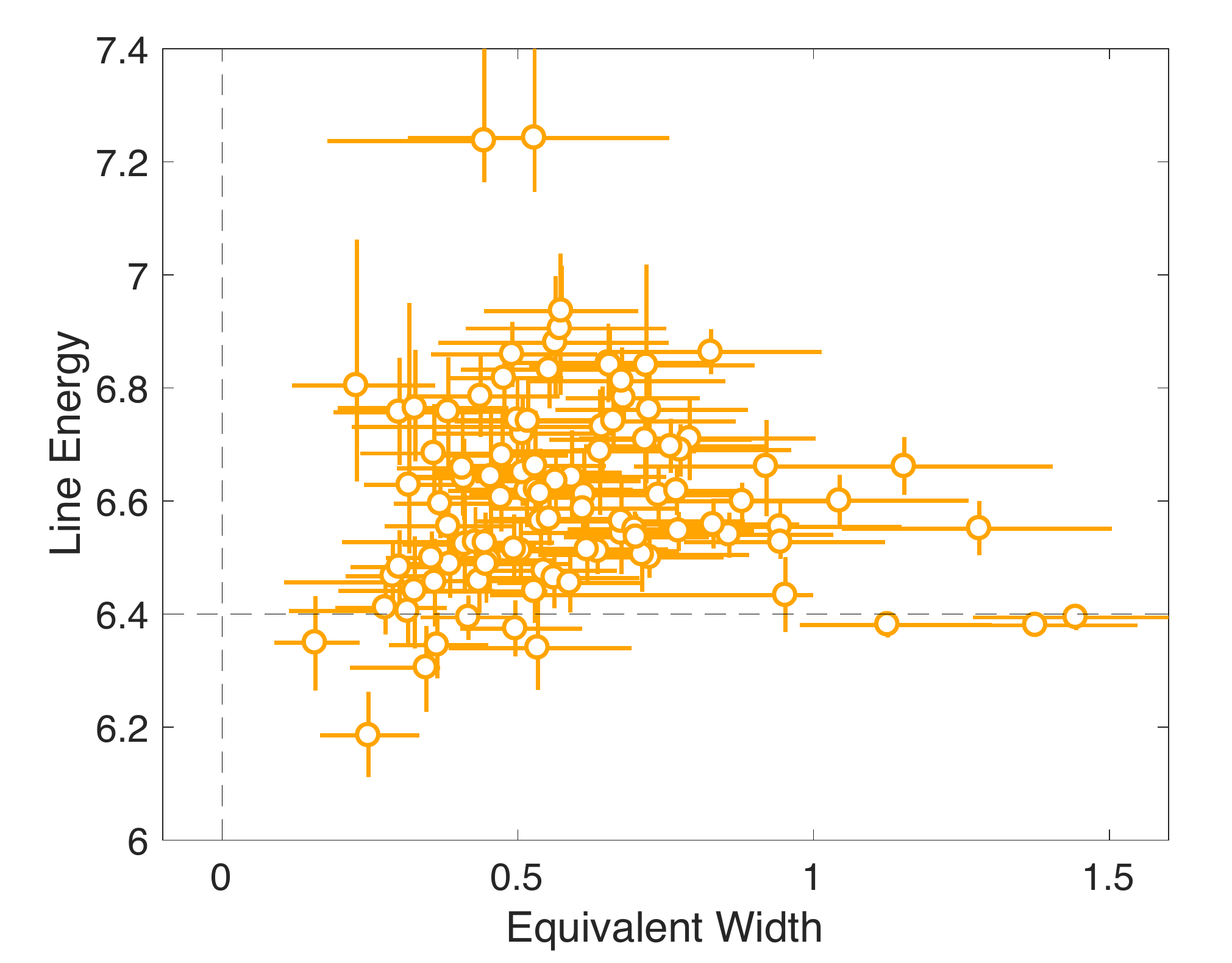}
\caption{Line energy versus line equivalent width for the observations that show the presence of an iron line. The black dashed line marks the 6.4 keV energy. }
\label{fig:eqwidth}
\end{figure}

As described in Sec. \ref{sec:analysis}, we have modelled features around 6.4~keV with a narrow Gaussian line. The resulting distribution of line energies is qualitatively a distribution peaked at 6.5~keV and slightly skewed towards the higher energies (see Fig. \ref{fig:corner}, panel o).  Figure \ref{fig:eqwidth} shows the line energy versus the line equivalent width: the line energy is generally shifted to energies higher than 6.4 keV (dashed line in the plot), but also - much more rarely - to energies lower then that, i.e. up to $\sim$~7.2~keV and down to $\sim$6.2~keV.

The spectral modelling of the iron line region is complicated by the presence of absorption/reflection edges and it is often difficult to determine what model best fits the data. Figure \ref{fig:spectra} (panel f) displays three cases where a line around 6.4 keV is required: the residuals in the iron line region highlight  the complexity of the spectra.
\citet{King2015} reported the presence of several emission lines in the iron region. In order of intensity, the \textit{Chandra} data show the following emission lines: iron-K$\alpha$ (by far strongest detected, indicating the presence of a large fraction of neutral iron), iron-K$\beta$, iron XXVI-Ly$\alpha$,  iron XXV recombination line, intercombination line and  forbidden line. In addition, they reported the presence of more lines in different regions of the X-ray energy spectrum, the strongest of which is the Si XIV Ly$\alpha$ line, at $\sim$2 keV.
In almost all spectra where an emission line was statistically required by our fits, we detected only one significant line at a time. The only exception is spectrum No. 4 (Obs. ID. 00031403038, snapshot 1, see Tab. \ref{tab:log}), where together with a line at $\sim$6.4 keV (that can be associated with a iron-K$\alpha$ line), we detected the hint of a line at $\sim$6.95 keV, that could be associated with the iron XXVI-Ly$\alpha$ line. 
Based on this assumption, a fit to this spectrum performed substituting the power law continuum and the Gaussian line with \textsc{xillver}\footnote{\textsc{Xillver} is a power law spectrum convolved with a reflection kernel. Reflection is modelled by solving radiation transfer on a plane-parallel, 1-dimensional slab with constant density (Garcia et al. 2010, 2013).} returns continuum parameters consistent with those obtained with the power law model: $\Gamma$ $\sim$ 1.5, \textrm{log}$\xi$ $\sim$ 1.5 (where $\xi$ is the ionization parameter, $\xi \equiv \textrm{L/(n r}^2)$), and iron abundance $\sim$4 with respect to the solar abundance, which suggests that mild ionization might be present at the time of spectrum No. 4. The iron abundance in the surface disk atmosphere of a torus-like geometry is expected to be higher than in the accretion flow as a whole. This is because certain ions are preferentially dragged upward owing to their magnetic moments, similarly to what happens in the solar corona where abundances differ from the solar photosphere \citep{Reynolds2012a}. 
Since no other spectra show evidence for a line beside the strongest detected line (that should be associated with the strongest line detected by \textit{Chandra}, therefore the iron-K$\alpha$ emission line), this result suggests that the ionization level should be lower in all the remaining \textit{Swift}/XRT spectra. 

In principle, however, when one single line is detected in the iron region, it could be either produced by neutral iron (resulting in a iron-K$\alpha$) or by iron in different ionization states. The strongest line produced by ionized iron is the iron XXVI-Ly$\alpha$ line, which should co-exist with the Si XIV Ly$\alpha$ line, produced with a flux significantly lower than that of the iron XXVI-Ly$\alpha$ line at Solar metallicities. The ratio between the Si XIV Ly$\alpha$ line and the iron XXVI-Ly$\alpha$ line is expected to be constant for a given plasma temperature. The two \textit{Chandra} observations return an average ratio $R_\textrm{Si/Fe} = 0.45$.  

The easiest interpretation would be that the detected line corresponds to a blue- or (seldom) red-shifted iron-K$\alpha$ line. However, we tested our data  in order to exclude that the detected line is instead to be associated with the iron XXVI-Ly$\alpha$ transition. Assuming that the line detected in the spectra is the iron XXVI-Ly$\alpha$ line, we estimated the expected flux of the Si XIV Ly$\alpha$ transition line using the ratio $R_\textrm{Si/Fe}$. Then we measured the upper limit of the Si XIV Ly$\alpha$ from each spectrum where an emission line is detected in the iron region and we compared it to the estimated Si XIV Ly$\alpha$ flux. We found that the only case where the upper limit to the measured Si XIV Ly$\alpha$ flux is consistent with the estimated Si XIV Ly$\alpha$ flux is that of spectrum No. 4. 

This confirms that the ionization is probably moderate and it does not produce significant features in the \textit{Swift}/XRT spectra. Hence, we conclude that the emission line we detect in a significant fraction of our spectra is indeed the iron-K$\alpha$ emission line, often affected by an energy shift.
We note, however, that assuming a constant ratio  $R_\textrm{Si/Fe}$ is not necessarily correct, especially considering the very variable environment that is the accretion flow around V404 Cyg. 
Additionally, we stress that we cannot exclude the possibility that lines from both neutral and ionized iron could be produced by different regions of the accretion flow. In this case more then one line could be present in a given spectrum, and each  would be affected by a different energy shift that could in principle push them together, making it impossible to resolve them with \textit{Swift}/XRT. Unfortunately, our data cannot be tested against this possibility.

From the instrumentation point of view, it must be noted that bright sources could easily fill the charge traps in \textit{Swift}/XRT (by so-called `sacrificial charge'), causing an energy shift in the spectrum of up to 50--100 eV (see http://www.swift.ac.uk/analysis/xrt/digest\_cal.php\#traps). Since we used the latest \textit{Swift}/XRT calibration files, this effect is probably limited to its minimum, but residual effects are still possible. This effect is particularly evident for very bright sources (count rate $>60\,\cts$ in 1 single CCD column). As noted before we detected emission lines around 6.4 keV only at fluxes strictly lower than 2$\times$10$^{-7}\,\ergcms$, which corresponds to count rates lower than 1 counts~s$^{-1}$~column$^{-1}$, therefore the aforementioned effect is not likely to be relevant. Furthermore, the maximum shift of the line energy that we observed was $\sim$ 0.5 keV and cannot be explained in terms of calibration issues.

\begin{figure*}
\centering\includegraphics[width=0.98\textwidth]{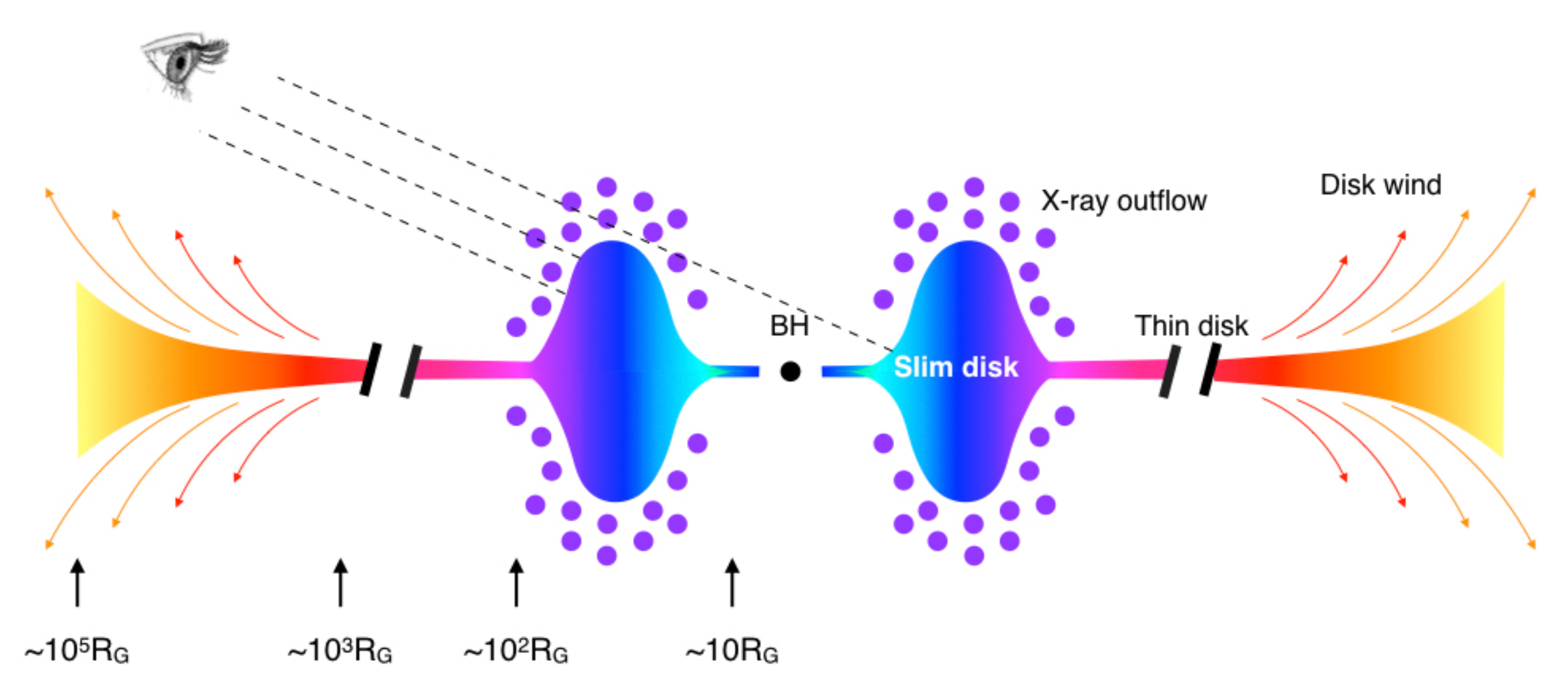}
\caption{Schematic view of V404 Cyg, mostly based on the findings reported in this work, in \citet{Motta2017} and \citet{Sanchez-Fernandez2017}, but also \citet{King2015} and \citet{Munoz-Darias2016}.
The inner region of the accretion flow is inflated and originates the clumpy outflow producing the heavy, non-uniform layer of absorbing material local to the source. The outer disk, instead, is responsible for the launch of the strong thermal wind seen at optical wavelengths. Note that \textit{slim disk} in fact refers to an accretion-flow that has a significant vertical extent compared to a \textit{thin disk}. Note that the X-axis of the figure is logarithmic. A colour version of this Figure is available on-line.}
\label{fig:sketch}
\end{figure*}


\section{Discussion}\label{sec:discussion}

We have analysed the \textit{Swift}/XRT data from the most active phase of the  outburst shown by the black hole binary V404 Cyg in June 2015. 
We performed time resolved spectroscopy on the source by extracting a total of 1054 spectra with varying exposures between 16 and a few hundred seconds. 
We fitted all the spectra with a phenomenological model constituted by a partially-covered, absorbed power law, to which we added a Gaussian line in emission. 
The results of the spectral fits revealed remarkable spectral variability on time scales down to a few tens of seconds, and variations in flux as large as 3 orders of magnitude in time--spans as short as a few minutes. 

The unique properties and behaviour shown by V404 Cyg during the June 2015 outburst can be explained by assuming that a heavy layer of non-uniform, high-column density material local to the source is affecting the observed emission, introducing the dramatic spectral and flux changes observed. This picture was already proposed by \cite{Motta2017} and is confirmed by our finding.
 
\subsection{The outflowing absorber}

\subsubsection{The absorber} 

The results of our spectral analysis show that spectral changes occurring on short time-scales (tens of seconds and likely less) can be ascribed to the variations of a layer of material covering the source. In particular, while changes in column density cause fast ($\sim$ minutes) but moderate flux changes, variations in the covering fraction of such an high-density absorber are responsible for the most dramatic flux and spectral changes observed during the outburst. What we observed is in agreement with what has been reported by e.g. \cite{Oosterbroek1996} and by \cite{Syunyaev1991} based on data collected with \textit{Ginga} and the Roentgen telescope, respectively, during the 1989 outburst, but also by \cite{Sanchez-Fernandez2017} and \cite{Motta2017}, based on \inte\ and simultaneous \textit{Swift} and \inte\ data, respectively, collected during the 2015 outburst. 

Our findings show that the absorber local to the source has a clear bi-modal behaviour: it either appears to be very dense (i.e. often Compton-thick, with average $N_\textrm{H}$ corresponding to $\tau \sim 1$) and highly inhomogeneous/clumpy, or tenuous and uniform. 
This could mean that the inhomogeneous high-density absorber and the uniform, tenuous one coexist at all times, even though our data are sensitive to the effects of the latter only when the high density, clumpy absorber is not present. Another possibility is that the clumpy absorber partially or completely rarefies at times into a much more uniform and thin layer of material, with density $\sim 10^{21}\,\textrm{cm}^{-2}$. However, this possibility seems less likely, given the fast transitions between high and low column density phases. 
Our data do not allow us to determine the relative position of these two absorbers with respect to the source (even though we can estimate the radial extent of the clumpy absorber, see below). Therefore, in the case of co-existence of the two absorbers, we cannot exclude that the uniform, thin absorber could sit further out with respect to the clumpy absorber, a possibility that is supported by the fact that the thin absorber does not appear to be ionised. Therefore, in principle, such thin absorber could even correspond to the neutral outflowing material launched as an optical wind at hundreds of thousands of $R_\textrm{g}$ from the central BH \citep{Munoz-Darias2016}. 

We note that the \textit{Swift}/XRT limited energy range imposes limits on our sensitivity to high values of column density, especially given the fact that a large fraction of the spectra in our sample require partial covering. Because of a lack of high-energy coverage, when it comes to the actual spectral fitting, our data would not allow us to distinguish between a model that requires only weak or even absent absorption, from a model that requires heavy absorption (with column density values above $\sim$10$^{23} \textrm{cm}^{-2}$), but also partial covering. In other words, we would be unable to detect the effects of a very high column density (which would be visible well above 10\, keV), and in the \textit{Swift}/XRT band-pass we would observe only the fraction of the emission that reaches the observer through the holes in the inhomogeneous absorber. The result would be a spectrum that appears, in such energy band, indistinguishable from a spectrum that is mildly absorbed, but with a much lower flux. 
For this reason, while we observed already fairly high column densities, we cannot exclude that the density of the absorber might have been at times much higher (i.e. in the Compton thick regime, $N_\textrm{H} > 10^{23}\, \textrm{cm}^{-2}$) than our measured values. For these reasons, the intrinsic luminosity could at times greatly differ (even by orders of magnitude) from the observed luminosity. This issue can be addressed by using \textit{Swift}/XRT plus a higher band-pass instrument, similarly to what had been done, for instance, by \cite{Motta2017}. In \cite{Motta2017} we used strictly simultaneous \textit{Swift}/XRT and \textit{INTEGRAL}/JEM-X and \textit{INTEGRAL}/IBIS-ISGRI data taken during a short-lived phase when the source was exceptionally showing little spectral variation. \textit{INTEGRAL}, \textit{Swift}/BAT and \textit{NuSTAR} data covering the hard X-rays were in fact collected during the 2015 outburst of V404 Cyg. However, time-resolved broad band spectroscopy is severely limited by the non-simultaneity of such data with those from \textit{Swift/XRT}. Furthermore, the short exposure times used to extract the XRT spectra would not allow us to obtain simultaneous spectra with enough S/N from either \textit{Swift}/BAT or \textit{INTEGRAL}/IBIS-ISGRI in most cases.

\subsubsection{The relation between flux and absorption}

Our results show that there is a clear link between the strongest flares seen by \textit{Swift} and the drop in column density. In particular, the strongest flares caught by \textit{Swift} seem to be related to sudden drops in the local column density and simultaneous disappearance of partial covering (see Sec. \ref{Sec:flux_evolution}). 
Such changes, where the source flux drops by almost three order of magnitude in minutes are too fast to be consistent with accretion-driven changes that typically follow  the viscous time-scale, and which would be larger than 10$^3$~s at 10 $R_\textrm{g}$ for a 10\, M$_{\odot}$ BH \citep{Frank2002}, and increasingly larger at larger radii. 
Comparable changes in flux, corresponding to an increase or drop in luminosity of a factor of five in a few seconds, have been observed in GRS 1915+105 \citep{Belloni1997}, and interpreted as a consequence of a thermal/viscous instability \citep{Lightman1974}. Such instability, however, is expected to produce a dramatic decrease in the soft emission (therefore a significant spectral hardening) as the flux drops. In V404 Cyg, instead, the decrease of the observed spectral slope  is most evident in correspondence of the strongest flares, while an increase of photon index follows as the flux drops again. 

Based on the above, the observed changes in the flux can hardly be explained as a consequence of an intrinsic change in the properties of the accretion flow. Instead, the covering/uncovering of a bright central source by clumps of matter can easily explain fast changes in luminosity. If this is the case, the super-Eddington flares observed in V404 Cyg would correspond to times where the actual emission produced by the source reaches the observer unabsorbed, while the lower flux phases would be explained in terms of partial or complete covering of the source.
Such a scenario, while seldom observed in BH binaries (see the case of V4641 Sgr,  \citealt{Revnivtsev2002} and SS 433, \citealt{Fabrika2004}), is often observed in obscured AGN, where the presence of a variable, patchy, neutral absorber (see, e.g., the case of NGC 4151, \citealt{Zdziarski2002a}, \citealt{Derosa2007} and NGC 1365, \citealt{Risaliti2005}) is thought to be responsible for most of the fast spectral and flux variability. A particularly interesting case is that of the Seyfert 1 galaxy NGC 5548 \citep{Mehdipour2016}, where changes in the covering fraction (ranging between 0.7 and 1) in an obscuring, weakly-ionised outflow close to the accretion disk is the primary and dominant cause of variability in the soft X-ray band.

\subsubsection{The relations between the X-ray and radio flux}

The most striking characteristic of both the radio and X-ray light curves is the very strong variability (see e.g. \citealt{Tetarenko2017}), characterised by hundreds of flares over a two-weeks period. Such extreme variability is probably revealed due to the relative proximity of the source and the exceptionally dense X-ray and radio coverage. At present there is no clear simple relation between peaks in the radio flux and the events in X-rays (i.e., in the accretion flow). This is most likely because \emph{(i)} the jet behaviour is more complex than previously anticipated (Fender et al. in prep), and \emph{(ii)} the effects of the strongly variable absorption described above makes it very difficult to recover the source intrinsic (i.e. accretion-driven) variations. 

The times when a few ejection events occurred  have so far been inferred based on the modelling\footnote{To date, the ejection times based on the detection of resolved ballistic ejecta have not been reported. }  of a short-exposure ($\sim$ 4 hours) set of radio and sub-millimetre observations taken with the Very Large Array, the Sub-millimetre Array and the James Clerk Maxwell Telescope (see \citealt{Tetarenko2017}), which unfortunately only marginally overlap with our \textit{Swift/XRT} data. The radio and sub-millimetre light curves display extraordinary flaring activity, that can be ascribed to eight different discrete bi-polar ejections (indicated by red arrows in Fig. \ref{fig:ISGRI}), of which only two occur when \swift/\textit{XRT} was observing the source (spectra 470 and 484). The results obtained by \citealt{Tetarenko2017} show that even an extensive multi-band modelling of radio and sub-mm light curves do not allow to draw a clear one-to-one association between ejection events and radio/X-ray flares.

Comparing the AMI radio light curve in Fig. \ref{fig:ISGRI}, we note that the largest flare observed in the hard X-rays (starting approximately at MJD 57199 and clearly visible in the \textit{INTEGRAL}/IBIS-ISGRI light curve, see Fig. \ref{fig:ISGRI}, top panel) is characterised by a higher amplitude and a longer duration with respect to all the other flares seen by \inte. Such an X-ray flare might be associated to a major radio flaring event, starting at $\approx$ MJD 57198 and likely peaking around MJD 57198.5.
\textit{Swift}/XRT data simultaneous with the \inte/IBIS-ISGRI data taken during such an X-ray flare (spectra 941-968 in Fig.  \ref{fig:ISGRI} and Fig. \ref{fig:parameters}) show that the source intrinsic flux is extremely high, while the local absorption has dropped to values consistent with zero (except in three spectra). While it might be possible that these phenomena are related and that a strong radio ejection associated with the major radio flare has ``cleared out'' the local absorber from the vicinity of the source, we would like to stress that the poor overlap between the \textit{Swift}/XRT and radio data makes any association between the X-ray and the radio behaviour very uncertain. While there is a significantly better overlap between the \textit{INTEGRAL} data and the \textit{AMI} data, a detailed comparison of such datasets is beyond the scope of this work and will be presented elsewhere (Fender et al. in prep).\\

\subsubsection{The emission lines and the absorber}

Our spectral analysis revealed the presence of an emission line in $\sim$10\% of our spectra. When detected, this line is narrow (i.e. not resolved by \textit{Swift}/XRT) and therefore likely produced  far away from the central black hole, as already suggested by \cite{King2015} and \cite{Motta2017} based on the data from the June 2015 outburst, and by \cite{Oosterbroek1997} based on the Ginga data from the 1989 outburst. 
Our findings show that the detected line corresponds to the iron-K$\alpha$ transition, and allow us to exclude its association with transitions of iron in different ionisation states. Our results also indicate that the ionization state of the material where the line originates is most of the time moderate to low  and thus it is not enough to produce significant effects in the \textit{Swift}/XRT spectra. 

The iron-K$\alpha$ line is observed only at phases of high column density, which suggests that the line might be produced in the same material that is causing absorption. This is particularly interesting considering the fact that the iron-K$\alpha$ line shows significant blue and sometimes red shifts (up to $7.2$~keV and down to $6.2$~keV, respectively). This is a clear indication of the presence of an outflow, where the blue shift corresponds to material launched in the direction of the observer and the red-shift to material launched away from it. 
In this context, the outflowing material would be therefore directly responsible for the dramatic, variable absorption observed in the data described above. 

Since the orbital (disk) inclination of the source is known ($i \approx 67^{\circ}$), 
the blue and red shifts of the line can be used to obtain an estimate of the outflow velocity and of the outflow launch angle with respect to the line of sight. For simplicity, we assume that the outflow is launched at a constant angle with respect to the disk plane, at a constant velocity, and that the maximum blue and red shifts of the line correspond to the maximum projected outflow velocities with respect to the line of sight towards and away from the observer, respectively. We obtained an upper limit to the outflow velocity of 0.1~c, slower than the outflow velocity of the ejecta observed in radio (Miller-Jones in prep.).
We also obtained an upper limit to the outflow launching angle of 35$^{\circ}$ from the disk plane. This implies that our line of sight passes straight through the slim disk/outflow, explaining the dramatic effects of the absorber on the source emission. 
The fact that the iron-K$\alpha$ line appears preferentially blue-shifted rather than red shifted is therefore a consequence of the relative inclination of the source with respect to the line of sight. 

Such a neutral, clumpy and variable outflow has never been observed with such clarity in any other BHB in outburst. The only BHB that has temporarily shown a limited increase in the local column density following a significant increase in the mass accretion rate is GRO J1655-40, which crossed the Ultra-Luminous state in both its 1995 and 2005 outbursts. During such high-luminosity events the source might have produced a neutral outflow from its inner accretion flow (\citealt{Kalemci2016}) that temporarily affected the source spectra with visible absorption features. 

\subsection{The source intrinsic emission}

\subsubsection{The slim disk}

In order to be able to launch a non-homogeneous outflow the accretion flow must be thick and sustained by strong radiation forces developed as a consequence of high (super-Eddington) accretion rates, forming a so-called \textit{slim disk} \citep{Abramowicz1988}. The outflow launch angle that we inferred above therefore sets an upper limit to the slim disk half-opening angle as well. A thick  flow half-opening angle of $<$ 35$^{\circ}$ is consistent with the half-opening angle found in MHD simulations of BHs accreting in a super-Eddington regime (e.g. \citealt{Takeuchi2013}), which can be as large as $\lesssim$80${^{\circ}}$ at extreme accretion rates.

The radiation pressure is thought becomes important in the inner region of the accretion flow, determining the structure of the slim disk. Assuming that the accretion rate is close to or larger than the Eddington accretion rate, radiation pressure is expected to exceed gas pressure for disk radii smaller than $\sim$10$^8$ cm (\citealt{Frank2002}, eq. 5.57) which corresponds to $\sim$10$^2$ $R_\textrm{g}$ for the case of V404 Cyg. This should roughly correspond to the radial extent of the slim disk (see Fig. \ref{fig:sketch}). Assuming that the outflow velocity is at least equal to the escape velocity from the BH at the clump distance, we can obtain an estimate of the clump distance from the BH, which is also of the order of 10$^2$ $R_\textrm{g}$. This supports the hypothesis that the clumpy outflow is indeed produced by the slim disk.

As the disk puffs up, it likely blocks a large fraction of the direct X-rays from the very central part of the accretion flow. However, as shown by \cite{Motta2017}, such radiation is scattered through the clumpy absorber surrounding the source. Therefore, the scattered radiation -- together with a small fraction of unabsorbed direct radiation -- might be able to illuminate the outer disk, possibly triggering the radiation or thermally driven X-ray winds detected in the X-rays \citep{King2015}, and launched at $\lesssim$ 10$^5$ $R_\textrm{g}$ from the central black hole.
At the same time, given the high accretion rates expected in the accretion flow, the outer walls of the thick disk are likely to become very UV bright \citep{Poutanen2007}, similarly to what is seen in certain ULXs observed at high inclination \citep{Kaaret2010}. Such strong UV radiation could illuminate the outer disk, perhaps contributing to the formation of the radiation or line driven neutral winds observed at optical wavelengths \citep{Munoz-Darias2016}, launched at $\sim$10$^5$ $R_\textrm{g}$ from the BH (see Fig. \ref{fig:sketch}).
In this scenario, a stratification in the ionization state of the outflow (higher towards the central part of the system and lower further out) would explain both the X-ray wind and the optical wind, with the ionized X-ray wind smoothly changing into the cooler optical wind for increasing launch radii.
The data currently do not allow us to exclude whether other sources of radiation -- such as hard emission originating a few to tens of $R_\textrm{g}$ above the accretion disk, e.g. in a jet -- could be responsible for, or contribute to, the outer disk illumination.

\subsubsection{The source spectrum}

The spectral slope versus flux changes observed in the \textit{Swift}/XRT data from V404 Cyg do not reflect the typical softer-when-brighter behaviour of other BHBs, where the increase in flux systematically corresponds to a constant or increasing photon index. Such characteristic trends are recognised to give rise to the well-known Q-shaped pattern in the HID \citep{Homan2001} of many BHBs. In V404 Cyg, instead, most of the increments in flux correspond to spectral hardening, where the photon index drops to values significantly lower than those expected for a Compton spectrum ($\Gamma$ $\sim$ 1.4) and sometimes even consistent with zero. 
Our results show that such extremely hard spectra are likely significantly affected by reflection/reprocessing, and could be at times explained by reflection/scattering dominated emission (see also \citealt{Motta2017}). 

The unusual relation between photon index and flux observed in V404 Cyg could be ascribed to the fact that the inner accretion flow does not match the standard Shakura-Sunyaev thin disk typically observed in BHBs \citep{Shakura1973}. As noted above, the inner accretion flow most likely takes the form of a slim disk -- whose spectrum is harder than that of a standard thin disk -- inserted in a peculiar framework where a cold, dense outflow is launched by the slim disk itself. 
It is particularly noteworthy that all the cases of spectral hardening accompanied by an increase in flux, also correspond to a drop in column density and/or decrease in partial covering of the absorber. 
This correlation suggests that the photon index resulting from our spectral fits might not reflect the intrinsic spectral slope of the X-ray source. Rather it describes a complex spectral shape resulting from the peculiar geometric configuration of the system, further complicated by the narrow energy range of the \textit{Swift}/XRT. The photon index values from our fits must therefore be interpreted as \emph{phenomenological/apparent} ones, and do not necessarily reflect the spectral slope of the intrinsic source spectrum. This might explain the difference between the fairly hard spectra seen from V404 Cyg and the very steep ($\Gamma \sim 3-5$) spectra seen from GRO J1655-40 during its hyper-soft state (see, e.g. \citealt{Uttley2015}, \citealt{Shidatsu2016}, \citealt{Neilsen2016}), which has been interpreted as the result of the emission from a slim disk, surrounded by a Compton-thick, fully ionized wind. It must be noted, however, that while the outflow from the inner disk of V404 Cyg is only mildly ionised (see Sec. \ref{Sec:line_dist} and \citealt{King2015}), that from GRO J1655-40 was, instead, completely ionized. Similarly, while the emission from V404 Cyg was at times completely or almost completely dominated by a strong scattering spectrum (see \citealt{Motta2017} and \citealt{Sanchez-Fernandez2017}), reflection was apparently absent in the hyper-soft spectra from GRO J1655-40. These fundamental differences between the two sources are probably at the base of their remarkably different observed spectra.

In the slim disk scenario applied to V404 Cyg, the intense, relatively soft X-ray emission produced in the innermost regions of the accretion flow (i.e. in the last few $R_\textrm{g}$, blue fraction of the accretion flow at $\sim$10 $R_\textrm{g}$ from the black hole in Fig. \ref{fig:sketch}), comes from a region relatively small compared to the inner regions of the slim disk (cyan fraction of the slim disk in Fig. \ref{fig:sketch}). The latter is expected to produce harder intrinsic emission and is responsible for significant radiation reprocessing, giving rise to very hard emission (see \citealt{Motta2017}). As a result, while the innermost regions of the accretion flow would remain almost always and almost completely hidden behind the inflated disk, the walls of the slim disk would be easily revealed to the observer any time clumps of dense matter move temporarily away from the line of sight.
This would result in a significant \emph{relative} increase of hard flux with respect to the soft flux, and therefore in an apparent hardening of the spectrum and in an overall increase in flux. 
This configuration would explain why some of the spectra of our sample show an extremely hard apparent photon index, and also why only very few of the spectra appear softer then $\Gamma \sim 2.4$. Similarly, this configuration also explains why our spectra never show the presence of a soft excess that could be interpreted as disk-blackbody emission, even when the local absorption is so low as to be negligible (see \citealt{Cabanac2009}). 
Past works \citep{Oosterbroek1996} reported the detection of a short-lived disk component in the \textit{Ginga} data collected during the 1989 outburst of V404 Cyg. However, as noted already by \cite{Motta2017}, it is possible that such soft emission is in reality related to the soft emission from a dust scattering halo forming around the source, which could not be disentangled from the source emission because of the much more limited spatial information from \textit{Ginga}/LAC.

\cite{Radhika2016} recently reported the detection of a fairly strong disk component in the same \textit{Swift} data-sets that we are considering here, at odds with what we report in this work. However, we note that the treatment of the dust scattering halo adopted by \cite{Radhika2016} differs from ours. We limited to the minimum the possible contamination of the source emission by shrinking the extraction region to a small circle when the dust scattering halo was affecting the source, and used a background spectrum of a clean sky region (see Appendix \ref{App:analysis}, Sec. \ref{sec:halo}). In contrast, \cite{Radhika2016} used the dust scattering halo spectra as background spectra where the halo was present around the source. This method could result in a over-correction or in an inadequate correction of the source spectrum, as the dust scattering halo brightness density is not radially uniform with respect to the central source (the dust scattering halo is formed by concentric rings with different intensities, see \citealt{Beardmore2016}) and could affect very differently the source and the sky regions around it. We also note that \cite{Radhika2016} claimed that partial covering in the absorber local to the source is only needed at the times where the iron-K-$\alpha$ line is visible in the spectra, while we found that partial covering is needed in a large fraction of our spectra. A more detailed comparison with these results is beyond the scope of this work, which is mostly focussed on the spectral variability related to absorption local to the source.

\section{Summary and conclusions}								
	
During its 2015 outburst the black hole binary V404 Cyg showed highly variable emission that reached extreme fluxes. 
We analysed the \textit{Swift}/XRT data from the most active phase of the outburst and performed time resolved spectroscopy. Our results show that: 

\begin{itemize}

\item The source is surrounded by a layer of high density, non-homogeneous, variable absorbing material local to it that heavily affects the source spectral shape and flux.

\item The largest variations in flux can be ascribed to dramatic changes in the absorber. In particular, the largest flares seen from the source correspond to significant drops in column density and covering of the absorber. 

\item A narrow iron-K$\alpha$ line is observed only in those spectra that show features caused by high column density material heavily absorbing the source emission.  

\item The iron line is significantly blue-shifted, and sometimes red-shifted, indicating the presence of a relatively fast neutral (or only mildly ionized) outflow. The characteristics of the iron line energy distribution  allow us to obtain an upper limit to the X-ray cold outflow launch angle ($\lesssim$35$^{\circ}$) and velocity ($v_\textrm{outflow} \lesssim 0.1$~c).

\item The photon index variations as a function of the source flux do not follow the expected softer-when-brighter trend typically observed in other BH binaries. The source shows instead significant spectral hardening for increasing fluxes and \textit{vice versa}. Such variations also appear to be correlated with the changes in the absorber. 

\item The data never showed the presence of soft emission that could be interpreted as a disk-blackbody spectrum, even at times where the absorber local to the source had negligible column density.

\end{itemize}

\noindent Based on these findings, we conclude that:

\begin{itemize}

\item The inner part of the accretion flow in V404 Cyg is likely a slim disk with half-opening angle $\sim$35$^{\circ}$, inflated because of strong radiative forces arising from the high, possibly (super-)Eddington accretion rates. Such a slim disk becomes therefore unstable and is able to launch a cold, clumpy outflow at a speed $\sim$0.1\,c, responsible both for the strong absorption and reflection features in the spectra, and whose variations introduce the remarkable spectral and flux changes observed from the source. This kind of outflow has never been observed before with such clarity in any accreting BHB.

\item The changes in the spectral slope of the \textit{Swift} spectra are a consequence of the changes in the variable absorber covering the source. The \textit{Swift} data alone do not allow us to distinguish between accretion-driven spectral and flux changes and the effects of covering-uncovering of the bright central source. Therefore we cannot exclude that part of the variability of the source is accretion-driven, similarly to what is seen in GRS 1915+105 or  other more well-behaved BH binaries (such as GX 339-4). 

\item Given the geometric configuration of the system, the intrinsic flux of the source could be at times or even constantly exceeding the Eddington luminosity, while being completely or partially obscured by non-homogeneous material with column densities in the Compton-thick regime or slightly below it. This implies that the observed flux might differ even by several orders of magnitude from the source intrinsic flux. 

\item V404 Cyg is able to produce simultaneously different types of outflows that are normally considered mutually exclusive in other black-hole binaries. Such outflows are complex, as they include: (i) a cold, dense and non-homogeneous outflow from the inner part of the accretion flow, observed in soft X-rays; (ii) a steady compact jet and relativistic ejections, both of which are observed in the radio; (iii) a wind from the outer disk, visible in X-rays through narrow line X-ray spectroscopy, and in the optical, where the high opening angle leads to extremely marked P-Cygni profiles.

\end{itemize}

\section*{Acknowledgements}

SEM acknowledges Jon Miller, Miguel Pereira-Santaella, James Matthews and Chris Done for useful discussions. SEM also acknowledges the Violette and Samuel Glasstone Research Fellowship programme and the UK Science and Technology Facilities Council (STFC) for financial support, and the European Space Astronomy Centre (ESAC) for hospitality.
JJEK acknowledges support from the Academy of Finland grants 268740 and 295114 and the ESA research fellowship programme. APB, JPO and KLP acknowledge support from the UK Space Agency.
SEM, JJEK and CSF acknowledge support from the ESAC Faculty. DA acknowledges support from the Royal Society. PAC is grateful for the financial support of an Emeritus Fellowship from the Leverhulme Trust. SRO is grateful for the financial support of an Early Career Fellowship from the Leverhulme Trust. MG acknowledges NWO, the Netherlands Organisation for Scientific Research, for financial support. 
 
\bigskip
\bibliographystyle{mn2e.bst}
\bibliography{biblio} 

\begin{thebibliography}{}

\bibitem[\protect\citeauthoryear{{Abramowicz}, {Czerny}, {Lasota} \&
  {Szuszkiewicz}}{{Abramowicz} et~al.}{1988}]{Abramowicz1988}
{Abramowicz} M.~A.,  {Czerny} B.,  {Lasota} J.~P.,    {Szuszkiewicz} E.,  1988,
  \apj, 332, 646

\bibitem[\protect\citeauthoryear{{Arnaud}}{{Arnaud}}{1996}]{Arnaud1996}
{Arnaud} K.~A.,  1996, in {G.~H.~Jacoby \& J.~Barnes} ed., Astronomical Data
  Analysis Software and Systems V Vol.~101 of Astronomical Society of the
  Pacific Conference Series, {XSPEC: The First Ten Years}.
pp 17--+

\bibitem[\protect\citeauthoryear{{Barthelmy}, {D'Ai}, {D'Avanzo}, {Krimm},
  {Lien}, {Marshall}, {Maselli} \& {Siegel}}{{Barthelmy}
  et~al.}{2015}]{Barthelmy2015}
{Barthelmy} S.~D.,  {D'Ai} A.,  {D'Avanzo} P.,  {Krimm} H.~A.,  {Lien} A.~Y.,
  {Marshall} F.~E.,  {Maselli} A.,    {Siegel} M.~H.,  2015, GRB Coordinates
  Network, 17929

\bibitem[\protect\citeauthoryear{{Beardmore}, {Altamirano}, {Kuulkers},
  {Motta}, {Osborne}, {Page}, {Sivakoff} \& {Vaughan}}{{Beardmore}
  et~al.}{2015}]{Beardmore2015}
{Beardmore} A.~P.,  {Altamirano} D.,  {Kuulkers} E.,  {Motta} S.~E.,  {Osborne}
  J.~P.,  {Page} K.~L.,  {Sivakoff} G.~R.,    {Vaughan} S.~A.,  2015, The
  Astronomer's Telegram, 7736, 1

\bibitem[\protect\citeauthoryear{{Beardmore}, {Willingale}, {Kuulkers},
  {Altamirano}, {Motta}, {Osborne}, {Page} \& {Sivakoff}}{{Beardmore}
  et~al.}{2016}]{Beardmore2016}
{Beardmore} A.~P.,  {Willingale} R.,  {Kuulkers} E.,  {Altamirano} D.,  {Motta}
  S.~E.,  {Osborne} J.~P.,  {Page} K.~L.,    {Sivakoff} G.~R.,  2016, \mnras,
  462, 1847

\bibitem[\protect\citeauthoryear{{Belloni}, {Mendez}, {King}, {van der Klis} \&
  {van Paradijs}}{{Belloni} et~al.}{1997}]{Belloni1997a}
{Belloni} T.,  {Mendez} M.,  {King} A.~R.,  {van der Klis} M.,    {van
  Paradijs} J.,  1997, \apjl, 479, L145+

\bibitem[\protect\citeauthoryear{{Belloni}, {van der Klis}, {Lewin}, {van
  Paradijs}, {Dotani}, {Mitsuda} \& {Miyamoto}}{{Belloni}
  et~al.}{1997}]{Belloni1997}
{Belloni} T.,  {van der Klis} M.,  {Lewin} W.~H.~G.,  {van Paradijs} J.,
  {Dotani} T.,  {Mitsuda} K.,    {Miyamoto} S.,  1997, \aap, 322, 857

\bibitem[\protect\citeauthoryear{{Belloni} \& {Motta}}{{Belloni} \&
  {Motta}}{2016}]{Belloni2016}
{Belloni} T.~M.,  {Motta} S.~E.,  2016, in {Bambi} C.,  ed., Astrophysics of
  Black Holes: From Fundamental Aspects to Latest Developments Vol.~440 of
  Astrophysics and Space Science Library, {Transient Black Hole Binaries}.
p.~61

\bibitem[\protect\citeauthoryear{{Bernardini}, {Russell}, {Shaw}, {Lewis},
  {Charles}, {Koljonen}, {Lasota} \& {Casares}}{{Bernardini}
  et~al.}{2016}]{Bernardini2016}
{Bernardini} F.,  {Russell} D.~M.,  {Shaw} A.~W.,  {Lewis} F.,  {Charles}
  P.~A.,  {Koljonen} K.~I.~I.,  {Lasota} J.~P.,    {Casares} J.,  2016, \apjl,
  818, L5

\bibitem[\protect\citeauthoryear{{Burrows}, {Hill}, {Nousek}, {Kennea},
  {Wells}, {Osborne}, {Abbey}, {Beardmore}, {Mukerjee}, {Short}, {Chincarini},
  {Campana}, {Citterio} \& {Moretti}}{{Burrows} et~al.}{2005}]{Burrows2005}
{Burrows} D.~N.,  {Hill} J.~E.,  {Nousek} J.~A.,  {Kennea} J.~A.,  {Wells} A.,
  {Osborne} J.~P.,  {Abbey} A.~F.,  {Beardmore} A.,  {Mukerjee} K.,  {Short}
  A.~D.~T.,  {Chincarini} G.,  {Campana} S.,  {Citterio} O.,    {Moretti}
  e.~a.,  2005, \ssr, 120, 165

\bibitem[\protect\citeauthoryear{{Cabanac}, {Fender}, {Dunn} \&
  {K{\"o}rding}}{{Cabanac} et~al.}{2009}]{Cabanac2009}
{Cabanac} C.,  {Fender} R.~P.,  {Dunn} R.~J.~H.,    {K{\"o}rding} E.~G.,  2009,
  \mnras, 396, 1415

\bibitem[\protect\citeauthoryear{{Casares}, {Charles} \& {Naylor}}{{Casares}
  et~al.}{1992}]{Casares1992}
{Casares} J.,  {Charles} P.~A.,    {Naylor} T.,  1992, \nat, 355, 614

\bibitem[\protect\citeauthoryear{{de Rosa}, {Piro}, {Perola}, {Capalbi},
  {Cappi}, {Grandi}, {Maraschi} \& {Petrucci}}{{de Rosa}
  et~al.}{2007}]{Derosa2007}
{de Rosa} A.,  {Piro} L.,  {Perola} G.~C.,  {Capalbi} M.,  {Cappi} M.,
  {Grandi} P.,  {Maraschi} L.,    {Petrucci} P.~O.,  2007, \aap, 463, 903

\bibitem[\protect\citeauthoryear{{Dunn}, {Fender}, {K{\"o}rding}, {Belloni} \&
  {Cabanac}}{{Dunn} et~al.}{2010}]{Dunn2010}
{Dunn} R.~J.~H.,  {Fender} R.~P.,  {K{\"o}rding} E.~G.,  {Belloni} T.,
  {Cabanac} C.,  2010, \mnras, 403, 61

\bibitem[\protect\citeauthoryear{{Fabrika}}{{Fabrika}}{2004}]{Fabrika2004}
{Fabrika} S.,  2004, Astrophysics and Space Physics Reviews, 12, 1

\bibitem[\protect\citeauthoryear{{Fender}, {Homan} \& {Belloni}}{{Fender}
  et~al.}{2009}]{Fender2009}
{Fender} R.~P.,  {Homan} J.,    {Belloni} T.~M.,  2009, \mnras, 396, 1370

\bibitem[\protect\citeauthoryear{{Frank}, {King} \& {Raine}}{{Frank}
  et~al.}{2002}]{Frank2002}
{Frank} J.,  {King} A.,    {Raine} D.~J.,  2002, {Accretion Power in
  Astrophysics: Third Edition}

\bibitem[\protect\citeauthoryear{{Gallo}, {Fender} \& {Hynes}}{{Gallo}
  et~al.}{2005}]{Gallo2005}
{Gallo} E.,  {Fender} R.~P.,    {Hynes} R.~I.,  2005, \mnras, 356, 1017

\bibitem[\protect\citeauthoryear{{Gallo}, {Fender} \& {Pooley}}{{Gallo}
  et~al.}{2003}]{Gallo2003}
{Gallo} E.,  {Fender} R.~P.,    {Pooley} G.~G.,  2003, \mnras, 344, 60

\bibitem[\protect\citeauthoryear{{Gandhi}, {Littlefair}, {Hardy}, {Dhillon},
  {Marsh}, {Shaw}, {Altamirano}, {Caballero-Garcia}, {Casares}, {Casella},
  {Castro-Tirado}, {Charles}, {Dallilar}, {Eikenberry} \& {Fender}}{{Gandhi}
  et~al.}{2016}]{Gandhi2016}
{Gandhi} P.,  {Littlefair} S.~P.,  {Hardy} L.~K.,  {Dhillon} V.~S.,  {Marsh}
  T.~R.,  {Shaw} A.~W.,  {Altamirano} D.,  {Caballero-Garcia} M.~D.,  {Casares}
  J.,  {Casella} P.,  {Castro-Tirado} A.~J.,  {Charles} P.~A.,  {Dallilar} Y.,
  {Eikenberry} S.,    {Fender} e.~a.,  2016, \mnras, 459, 554

\bibitem[\protect\citeauthoryear{{Heinz}, {Corrales}, {Smith}, {Brandt},
  {Jonker}, {Plotkin} \& {Neilsen}}{{Heinz} et~al.}{2016}]{Heinz2016}
{Heinz} S.,  {Corrales} L.,  {Smith} R.,  {Brandt} W.~N.,  {Jonker} P.~G.,
  {Plotkin} R.~M.,    {Neilsen} J.,  2016, \apj, 825, 15

\bibitem[\protect\citeauthoryear{{Homan}, {Wijnands}, {van der Klis},
  {Belloni}, {van Paradijs}, {Klein-Wolt}, {Fender} \& {M{\'e}ndez}}{{Homan}
  et~al.}{2001}]{Homan2001}
{Homan} J.,  {Wijnands} R.,  {van der Klis} M.,  {Belloni} T.,  {van Paradijs}
  J.,  {Klein-Wolt} M.,  {Fender} R.,    {M{\'e}ndez} M.,  2001, \apjs, 132,
  377

\bibitem[\protect\citeauthoryear{{Hynes}, {Bradley}, {Rupen}, {Gallo},
  {Fender}, {Casares} \& {Zurita}}{{Hynes} et~al.}{2009}]{Hynes2009}
{Hynes} R.~I.,  {Bradley} C.~K.,  {Rupen} M.,  {Gallo} E.,  {Fender} R.~P.,
  {Casares} J.,    {Zurita} C.,  2009, \mnras, 399, 2239

\bibitem[\protect\citeauthoryear{{Kaaret}, {Feng}, {Wong} \& {Tao}}{{Kaaret}
  et~al.}{2010}]{Kaaret2010}
{Kaaret} P.,  {Feng} H.,  {Wong} D.~S.,    {Tao} L.,  2010, \apjl, 714, L167

\bibitem[\protect\citeauthoryear{{Kalemci}, {Begelman}, {Maccarone}, {Din{\c
  c}er}, {Russell}, {Bailyn} \& {Tomsick}}{{Kalemci}
  et~al.}{2016}]{Kalemci2016}
{Kalemci} E.,  {Begelman} M.~C.,  {Maccarone} T.~J.,  {Din{\c c}er} T.,
  {Russell} T.~D.,  {Bailyn} C.,    {Tomsick} J.~A.,  2016, \mnras, 463, 615

\bibitem[\protect\citeauthoryear{{Kimura}, {Isogai}, {Kato}, {Ueda},
  {Nakahira}, {Shidatsu}, {Enoto}, {Hori}, {Nogami}, {Littlefield}, {Ishioka},
  {Chen}, {King}, {Wen}, {Wang}, {Lehner}, {Schwamb} \& {Wang} J.-H.}{{Kimura}
  et~al.}{2016}]{Kimura2016}
{Kimura} M.,  {Isogai} K.,  {Kato} T.,  {Ueda} Y.,  {Nakahira} S.,  {Shidatsu}
  M.,  {Enoto} T.,  {Hori} T.,  {Nogami} D.,  {Littlefield} C.,  {Ishioka} R.,
  {Chen} Y.-T.,  {King} S.-K.,  {Wen} C.-Y.,  {Wang} S.-Y.,  {Lehner} M.~J.,
  {Schwamb} M.~E.,    {Wang} J.-H. e.~a.,  2016, \nat, 529, 54

\bibitem[\protect\citeauthoryear{{King}, {Miller}, {Raymond}, {Reynolds} \&
  {Morningstar}}{{King} et~al.}{2015}]{King2015}
{King} A.~L.,  {Miller} J.~M.,  {Raymond} J.,  {Reynolds} M.~T.,
  {Morningstar} W.,  2015, \apjl, 813, L37

\bibitem[\protect\citeauthoryear{{Kuulkers}}{{Kuulkers}}{2015}]{Kuulkers2015c}
{Kuulkers} E.,  2015, The Astronomer's Telegram, 7758

\bibitem[\protect\citeauthoryear{{Kuulkers}, {Motta}, {Kajava}, {Homan},
  {Fender} \& {Jonker}}{{Kuulkers} et~al.}{2015}]{Kuulkers2015}
{Kuulkers} E.,  {Motta} S.,  {Kajava} J.,  {Homan} J.,  {Fender} R.,
  {Jonker} P.,  2015, The Astronomer's Telegram, 7647, 1

\bibitem[\protect\citeauthoryear{{Liddle}}{{Liddle}}{2007}]{Liddle2007}
{Liddle} A.~R.,  2007, \mnras, 377, L74

\bibitem[\protect\citeauthoryear{{Lightman} \& {Eardley}}{{Lightman} \&
  {Eardley}}{1974}]{Lightman1974}
{Lightman} A.~P.,  {Eardley} D.~M.,  1974, \apjl, 187, L1

\bibitem[\protect\citeauthoryear{{Makino}, {Wagner}, {Starrfield}, {Buie},
  {Bond}, {Johnson}, {Harrison} \& {Gehrz}}{{Makino} et~al.}{1989}]{Makino1989}
{Makino} F.,  {Wagner} R.~M.,  {Starrfield} S.,  {Buie} M.~W.,  {Bond} H.~E.,
  {Johnson} J.,  {Harrison} T.,    {Gehrz} R.~D.,  1989, \iaucirc, 4786, 1

\bibitem[\protect\citeauthoryear{{Margon}}{{Margon}}{1984}]{Margon1984}
{Margon} B.,  1984, \araa, 22, 507

\bibitem[\protect\citeauthoryear{{Markoff}}{{Markoff}}{2010}]{Markoff2010}
{Markoff} S.,  2010, in {T.~Belloni} ed., Lecture Notes in Physics, Berlin
  Springer Verlag Vol.~794 of Lecture Notes in Physics, Berlin Springer Verlag,
  {From Multiwavelength to Mass Scaling: Accretion and Ejection in Microquasars
  and AGN}.
pp 143--+

\bibitem[\protect\citeauthoryear{{McClintock} \& {Remillard}}{{McClintock} \&
  {Remillard}}{2006}]{McClintock2006}
{McClintock} J.~E.,  {Remillard} R.~A.,  2006, pp 157--213

\bibitem[\protect\citeauthoryear{{Mehdipour}, {Kaastra}, {Kriss}, {Cappi},
  {Petrucci}, {De Marco}, {Ponti}, {Steenbrugge}, {Behar}, {Bianchi} \&
  {Branduardi-Raymont}}{{Mehdipour} et~al.}{2016}]{Mehdipour2016}
{Mehdipour} M.,  {Kaastra} J.~S.,  {Kriss} G.~A.,  {Cappi} M.,  {Petrucci}
  P.-O.,  {De Marco} B.,  {Ponti} G.,  {Steenbrugge} K.~C.,  {Behar} E.,
  {Bianchi} S.,    {Branduardi-Raymont} G. e.~a.,  2016, \aap, 588, A139

\bibitem[\protect\citeauthoryear{{Miller-Jones}, {Jonker}, {Nelemans},
  {Portegies Zwart}, {Dhawan}, {Brisken}, {Gallo} \& {Rupen}}{{Miller-Jones}
  et~al.}{2009}]{Miller-Jones2009}
{Miller-Jones} J.~C.~A.,  {Jonker} P.~G.,  {Nelemans} G.,  {Portegies Zwart}
  S.,  {Dhawan} V.,  {Brisken} W.,  {Gallo} E.,    {Rupen} M.~P.,  2009,
  \mnras, 394, 1440

\bibitem[\protect\citeauthoryear{{Motta}, {Homan}, {Mu{\~n}oz Darias},
  {Casella}, {Belloni}, {Hiemstra} \& {M{\'e}ndez}}{{Motta}
  et~al.}{2012}]{Motta2012}
{Motta} S.,  {Homan} J.,  {Mu{\~n}oz Darias} T.,  {Casella} P.,  {Belloni}
  T.~M.,  {Hiemstra} B.,    {M{\'e}ndez} M.,  2012, \mnras, 427, 595

\bibitem[\protect\citeauthoryear{{Motta}, {Kajava},
  {S{\'a}nchez-Fern{\'a}ndez}, {Giustini} \& {Kuulkers}}{{Motta}
  et~al.}{2017}]{Motta2017}
{Motta} S.~E.,  {Kajava} J.~J.~E.,  {S{\'a}nchez-Fern{\'a}ndez} C.,  {Giustini}
  M.,    {Kuulkers} E.,  2017, \mnras, 468, 981

\bibitem[\protect\citeauthoryear{{Mu{\~n}oz-Darias}, {Casares}, {Mata
  S{\'a}nchez}, {Fender}, {Armas Padilla}, {Linares}, {Ponti}, {Charles},
  {Mooley} \& {Rodriguez}}{{Mu{\~n}oz-Darias} et~al.}{2016}]{Munoz-Darias2016}
{Mu{\~n}oz-Darias} T.,  {Casares} J.,  {Mata S{\'a}nchez} D.,  {Fender} R.~P.,
  {Armas Padilla} M.,  {Linares} M.,  {Ponti} G.,  {Charles} P.~A.,  {Mooley}
  K.~P.,    {Rodriguez} J.,  2016, \nat, 534, 75

\bibitem[\protect\citeauthoryear{{Neilsen} \& {Lee}}{{Neilsen} \&
  {Lee}}{2009}]{Neilsen2009}
{Neilsen} J.,  {Lee} J.~C.,  2009, \nat, 458, 481

\bibitem[\protect\citeauthoryear{{Neilsen}, {Rahoui}, {Homan} \&
  {Buxton}}{{Neilsen} et~al.}{2016}]{Neilsen2016}
{Neilsen} J.,  {Rahoui} F.,  {Homan} J.,    {Buxton} M.,  2016, \apj, 822, 20

\bibitem[\protect\citeauthoryear{{Oosterbroek}, {van der Klis}, {van Paradijs},
  {Vaughan}, {Rutledge}, {Lewin}, {Tanaka}, {Nagase}, {Dotani}, {Mitsuda} \&
  {Miyamoto}}{{Oosterbroek} et~al.}{1997}]{Oosterbroek1997}
{Oosterbroek} T.,  {van der Klis} M.,  {van Paradijs} J.,  {Vaughan} B.,
  {Rutledge} R.,  {Lewin} W.~H.~G.,  {Tanaka} Y.,  {Nagase} F.,  {Dotani} T.,
  {Mitsuda} K.,    {Miyamoto} S.,  1997, \aap, 321, 776

\bibitem[\protect\citeauthoryear{{Oosterbroek}, {van der Klis}, {Vaughan}, {van
  Paradijs}, {Rutledge}, {Lewin}, {Tanaka}, {Nagase}, {Dotani}, {Mitsuda} \&
  {Yoshida}}{{Oosterbroek} et~al.}{1996}]{Oosterbroek1996}
{Oosterbroek} T.,  {van der Klis} M.,  {Vaughan} B.,  {van Paradijs} J.,
  {Rutledge} R.,  {Lewin} W.~H.~G.,  {Tanaka} Y.,  {Nagase} F.,  {Dotani} T.,
  {Mitsuda} K.,    {Yoshida} K.,  1996, \aap, 309, 781

\bibitem[\protect\citeauthoryear{{Plant}, {Fender}, {Ponti}, {Mu{\~n}oz-Darias}
  \& {Coriat}}{{Plant} et~al.}{2014}]{Plant2014}
{Plant} D.~S.,  {Fender} R.~P.,  {Ponti} G.,  {Mu{\~n}oz-Darias} T.,
  {Coriat} M.,  2014, \mnras, 442, 1767

\bibitem[\protect\citeauthoryear{{Ponti}, {Fender}, {Begelman}, {Dunn},
  {Neilsen} \& {Coriat}}{{Ponti} et~al.}{2012}]{Ponti2012}
{Ponti} G.,  {Fender} R.~P.,  {Begelman} M.~C.,  {Dunn} R.~J.~H.,  {Neilsen}
  J.,    {Coriat} M.,  2012, \mnras, 422, L11

\bibitem[\protect\citeauthoryear{{Poutanen}, {Lipunova}, {Fabrika}, {Butkevich}
  \& {Abolmasov}}{{Poutanen} et~al.}{2007}]{Poutanen2007}
{Poutanen} J.,  {Lipunova} G.,  {Fabrika} S.,  {Butkevich} A.~G.,
  {Abolmasov} P.,  2007, \mnras, 377, 1187

\bibitem[\protect\citeauthoryear{{Poutanen} \& {Veledina}}{{Poutanen} \&
  {Veledina}}{2014}]{Poutanen2014}
{Poutanen} J.,  {Veledina} A.,  2014, \ssr, 183, 61

\bibitem[\protect\citeauthoryear{{Radhika}, {Nandi}, {Agrawal} \&
  {Mandal}}{{Radhika} et~al.}{2016}]{Radhika2016}
{Radhika} D.,  {Nandi} A.,  {Agrawal} V.~K.,    {Mandal} S.,  2016, \mnras,
  462, 1834

\bibitem[\protect\citeauthoryear{{Revnivtsev}, {Gilfanov}, {Churazov} \&
  {Sunyaev}}{{Revnivtsev} et~al.}{2002}]{Revnivtsev2002}
{Revnivtsev} M.,  {Gilfanov} M.,  {Churazov} E.,    {Sunyaev} R.,  2002, \aap,
  391, 1013

\bibitem[\protect\citeauthoryear{{Reynolds}}{{Reynolds}}{2012}]{Reynolds2012a}
{Reynolds} C.~S.,  2012, \apjl, 759, L15

\bibitem[\protect\citeauthoryear{{Reynolds}}{{Reynolds}}{2014}]{Reynolds2014}
{Reynolds} C.~S.,  2014, \ssr, 183, 277

\bibitem[\protect\citeauthoryear{{Risaliti}, {Elvis}, {Fabbiano}, {Baldi} \&
  {Zezas}}{{Risaliti} et~al.}{2005}]{Risaliti2005}
{Risaliti} G.,  {Elvis} M.,  {Fabbiano} G.,  {Baldi} A.,    {Zezas} A.,  2005,
  \apjl, 623, L93

\bibitem[\protect\citeauthoryear{{Rodriguez}, {Cadolle Bel},
  {Alfonso-Garz{\'o}n}, {Siegert}, {Zhang}, {Grinberg}, {Savchenko}, {Tomsick},
  {Chenevez}, {Clavel}, {Corbel}, {Diehl}, {Domingo}, {Gouiff{\`e}s}, {Greiner}
  \& {Krause}}{{Rodriguez} et~al.}{2015}]{Rodriguez2015}
{Rodriguez} J.,  {Cadolle Bel} M.,  {Alfonso-Garz{\'o}n} J.,  {Siegert} T.,
  {Zhang} X.-L.,  {Grinberg} V.,  {Savchenko} V.,  {Tomsick} J.~A.,  {Chenevez}
  J.,  {Clavel} M.,  {Corbel} S.,  {Diehl} R.,  {Domingo} A.,  {Gouiff{\`e}s}
  C.,  {Greiner} J.,    {Krause} M.~G.~H. e.~a.,  2015, \aap, 581, L9

\bibitem[\protect\citeauthoryear{{S{\'a}nchez-Fern{\'a}ndez}, {Kajava}, {Motta}
  \& {Kuulkers}}{{S{\'a}nchez-Fern{\'a}ndez}
  et~al.}{2017}]{Sanchez-Fernandez2017}
{S{\'a}nchez-Fern{\'a}ndez} C.,  {Kajava} J.~J.~E.,  {Motta} S.~E.,
  {Kuulkers} E.,  2017, \aap, 602, A40

\bibitem[\protect\citeauthoryear{Schwarz et~al.,}{Schwarz
  et~al.}{1978}]{Schwarz1978}
Schwarz G.,  et~al., 1978, The annals of statistics, 6, 461

\bibitem[\protect\citeauthoryear{{Shakura} \& {Sunyaev}}{{Shakura} \&
  {Sunyaev}}{1973}]{Shakura1973}
{Shakura} N.~I.,  {Sunyaev} R.~A.,  1973, \aap, 24, 337

\bibitem[\protect\citeauthoryear{{Shidatsu}, {Done} \& {Ueda}}{{Shidatsu}
  et~al.}{2016}]{Shidatsu2016}
{Shidatsu} M.,  {Done} C.,    {Ueda} Y.,  2016, \apj, 823, 159

\bibitem[\protect\citeauthoryear{{Sivakoff}, {Bahramian}, {Altamirano},
  {Beardmore}, {Kuulkers} \& {Motta}}{{Sivakoff} et~al.}{2015}]{Sivakoff2015}
{Sivakoff} G.~R.,  {Bahramian} A.,  {Altamirano} D.,  {Beardmore} A.~P.,
  {Kuulkers} E.,    {Motta} S.,  2015, The Astronomer's Telegram, 7959, 1

\bibitem[\protect\citeauthoryear{{Syunyaev}, {Kaniovskii}, {Efremov}, {Arefev},
  {Borozdin}, {Gilfanov}, {Churazov}, {Kuznetsov}, {Melioranskii}, {Yamburenko}
  \& et al.}{{Syunyaev} et~al.}{1991}]{Syunyaev1991}
{Syunyaev} R.~A.,  {Kaniovskii} A.~S.,  {Efremov} V.~V.,  {Arefev} V.~A.,
  {Borozdin} K.~N.,  {Gilfanov} M.~R.,  {Churazov} E.~M.,  {Kuznetsov} A.~V.,
  {Melioranskii} A.~S.,  {Yamburenko} N.~S.,    et al. P.,  1991, Soviet
  Astronomy Letters, 17, 123

\bibitem[\protect\citeauthoryear{{Takeuchi}, {Ohsuga} \&
  {Mineshige}}{{Takeuchi} et~al.}{2013}]{Takeuchi2013}
{Takeuchi} S.,  {Ohsuga} K.,    {Mineshige} S.,  2013, \pasj, 65

\bibitem[\protect\citeauthoryear{{Tetarenko}, {Sivakoff}, {Miller-Jones},
  {Rosolowsky}, {Petitpas}, {Gurwell}, {Wouterloot}, {Fender}, {Heinz},
  {Maitra}, {Markoff}, {Migliari}, {Rupen}, {Rushton}, {Russell}, {Russell} \&
  {Sarazin}}{{Tetarenko} et~al.}{2017}]{Tetarenko2017}
{Tetarenko} A.~J.,  {Sivakoff} G.~R.,  {Miller-Jones} J.~C.~A.,  {Rosolowsky}
  E.~W.,  {Petitpas} G.,  {Gurwell} M.,  {Wouterloot} J.,  {Fender} R.,
  {Heinz} S.,  {Maitra} D.,  {Markoff} S.~B.,  {Migliari} S.,  {Rupen} M.~P.,
  {Rushton} A.~P.,  {Russell} D.~M.,  {Russell} T.~D.,    {Sarazin} C.~L.,
  2017, \mnras, 469, 3141

\bibitem[\protect\citeauthoryear{{Uttley} \& {Klein-Wolt}}{{Uttley} \&
  {Klein-Wolt}}{2015}]{Uttley2015}
{Uttley} P.,  {Klein-Wolt} M.,  2015, \mnras, 451, 475

\bibitem[\protect\citeauthoryear{{Valencic} \& {Smith}}{{Valencic} \&
  {Smith}}{2015}]{Valencic2015}
{Valencic} L.~A.,  {Smith} R.~K.,  2015, \apj, 809, 66

\bibitem[\protect\citeauthoryear{{Vasilopoulos} \&
  {Petropoulou}}{{Vasilopoulos} \& {Petropoulou}}{2016}]{Vasilopoulos2016}
{Vasilopoulos} G.,  {Petropoulou} M.,  2016, \mnras, 455, 4426

\bibitem[\protect\citeauthoryear{{Verner}, {Ferland}, {Korista} \&
  {Yakovlev}}{{Verner} et~al.}{1996}]{Verner1996}
{Verner} D.~A.,  {Ferland} G.~J.,  {Korista} K.~T.,    {Yakovlev} D.~G.,  1996,
  \apj, 465, 487

\bibitem[\protect\citeauthoryear{{Walton}, {Mooley}, {King}, {Tomsick},
  {Miller}, {Dauser}, {Garc{\'{\i}}a}, {Bachetti}, {Brightman}, {Fabian},
  {Forster} \& et al.}{{Walton} et~al.}{2017}]{Walton2017}
{Walton} D.~J.,  {Mooley} K.,  {King} A.~L.,  {Tomsick} J.~A.,  {Miller} J.~M.,
   {Dauser} T.,  {Garc{\'{\i}}a} J.~A.,  {Bachetti} M.,  {Brightman} M.,
  {Fabian} A.~C.,  {Forster} K.,    et al. F.,  2017, \apj, 839, 110

\bibitem[\protect\citeauthoryear{{Wijnands}, {Degenaar}, {Armas Padilla},
  {Altamirano}, {Cavecchi}, {Linares}, {Bahramian} \& {Heinke}}{{Wijnands}
  et~al.}{2015}]{Wijnands2015}
{Wijnands} R.,  {Degenaar} N.,  {Armas Padilla} M.,  {Altamirano} D.,
  {Cavecchi} Y.,  {Linares} M.,  {Bahramian} A.,    {Heinke} C.~O.,  2015,
  \mnras, 454, 1371

\bibitem[\protect\citeauthoryear{{Wilms}, {Allen} \& {McCray}}{{Wilms}
  et~al.}{2000}]{Wilms2000}
{Wilms} J.,  {Allen} A.,    {McCray} R.,  2000, \apj, 542, 914

\bibitem[\protect\citeauthoryear{{Zdziarski}, {Leighly}, {Matsuoka}, {Cappi} \&
  {Mihara}}{{Zdziarski} et~al.}{2002}]{Zdziarski2002a}
{Zdziarski} A.~A.,  {Leighly} K.~M.,  {Matsuoka} M.,  {Cappi} M.,    {Mihara}
  T.,  2002, \apj, 573, 505

\bibitem[\protect\citeauthoryear{{Zdziarski}, {Lubi{\'n}ski} \&
  {Smith}}{{Zdziarski} et~al.}{1999}]{Zdziarski1999}
{Zdziarski} A.~A.,  {Lubi{\'n}ski} P.,    {Smith} D.~A.,  1999, \mnras, 303,
  L11

\bibitem[\protect\citeauthoryear{{{\.Z}ycki}, {Done} \& {Smith}}{{{\.Z}ycki}
  et~al.}{1999}]{Zycki1999}
{{\.Z}ycki} P.~T.,  {Done} C.,    {Smith} D.~A.,  1999, \mnras, 309, 561

\end{thebibliography}

\appendix

\section{Data reduction and treatment of the dust scattering halo}\label{App:analysis}

\subsection{Data reduction: definition of the extraction regions.}\label{sec:datared}

Given the extreme brightness of the source, the XRT data are often significantly affected by photon pile-up (corresponding to a nominal count rate threshold of about $150\,\cts$), which is known to induce distortion of the XRT spectral response (see http://www.swift.ac.uk/analysis/xrt/pileup.php). 
High photo-electric absorption can also cause distortion of the spectral shape at low energies (i.e. below 1 keV, see http://www.swift.ac.uk/analysis/xrt/digest\_cal.php). 
For these reasons we treated our data as follows.

First, following the XRT reduction threads (http://www.swift.ac.uk/analysis/xrt/\#abs) we extracted only grade 0 events, which helps minimise the effects of pile-up when the source is bright ($> 150\,\cts$, see http://www.swift.ac.uk/xrt\_curves/cppdocs.php) and reduces the spectral distortion encountered in WT mode below 1.0 keV when the spectra are highly absorbed. Furthermore, we ignored data below 0.6 keV as at such low energies the spectra can be dominated by strong redistribution effects associated with the WT readout process, as well as trailing charge released from deep charge traps in the CCD on time-scales comparable to the WT readout time, which results in additional low energy events.

We extracted events in circular regions centred at the source position, with variable inner radius to mitigate pile-up, and outer radius fixed at 10, 20 or 30 pixels to adapt the photons extraction to the source properties.
In order to reduce the effects of pile-up to the minimum, we varied the inner radius of the extraction region until the spectral shape was no longer varying as a function of the inner radius and the count rate was lower than $\sim$150~counts~s$^{-1}$.
The outer radius had been determined based on the brightness of the source and depending on the presence of a dust scattering halo around it, which could contaminate the source emission (see Sec. \ref{sec:halo}). In general we used a 20 pixels outer radius extraction region. However, when the source was very bright, we extracted the source photons in a annular region with outer radius 30 pixels (see http://www.swift.ac.uk/analysis/xrt/spectra.php). Instead, in the cases when the presence of a dust scattering halo was likely contaminating the source (see Sec. \ref{sec:halo} for details), we shrank the outer radius of the extraction region to 10 pixels, in order to minimize the halo contamination.  In Tab. \ref{tab:log} we reported, for each observation/snapshot, the outer radius of the extraction region. 

We divided each event file thus obtained into time segments of variable duration. Then, we determined the duration of each segment by requiring that each segment contained about 1600 counts after mitigating the pile-up. We set the minimum exposure to 16s, while the maximum exposure we obtained was of 224s.
These requirements ensure that all the spectra have roughly constant overall S/N and that  they will be well described by the XRT response calibration files. This choice has been made in order to obtain a good compromise between S/N and sensitivity to the spectral variations -- that can occur on time-scales of seconds (see \citealt{Oosterbroek1996} and \citealt{Zycki1999}) -- and guarantees good sensitivity to fairly fast spectral changes, without sacrificing S/N. 

\subsection{Dust scattering halo contamination effects}\label{sec:halo}

\cite{Vasilopoulos2016}, \cite{Heinz2016} and \cite{Beardmore2016} reported the presence of a dust scattering halo in the regions close to the source during most of the June 2016 outburst. As shown by these authors, this component is variable on a time-scale significantly longer than the one over which the source itself varies, and can be considered constant during the average exposure of one \textit{Swift}/XRT snapshot (about 1ks), and hence also during the spectra of our sample. 

\begin{figure*}
\centering
\includegraphics[width=0.45\textwidth]{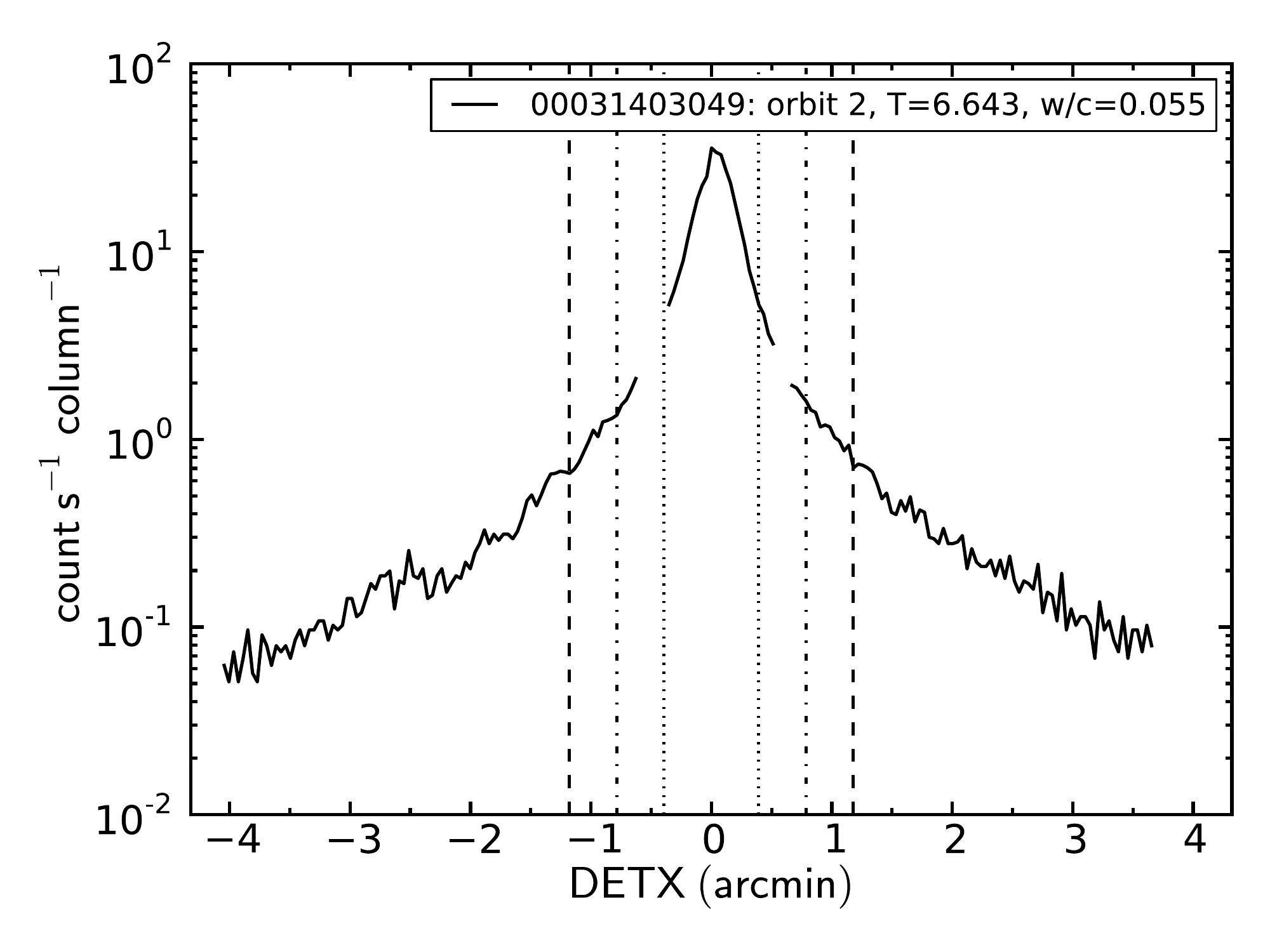}
\includegraphics[width=0.45\textwidth]{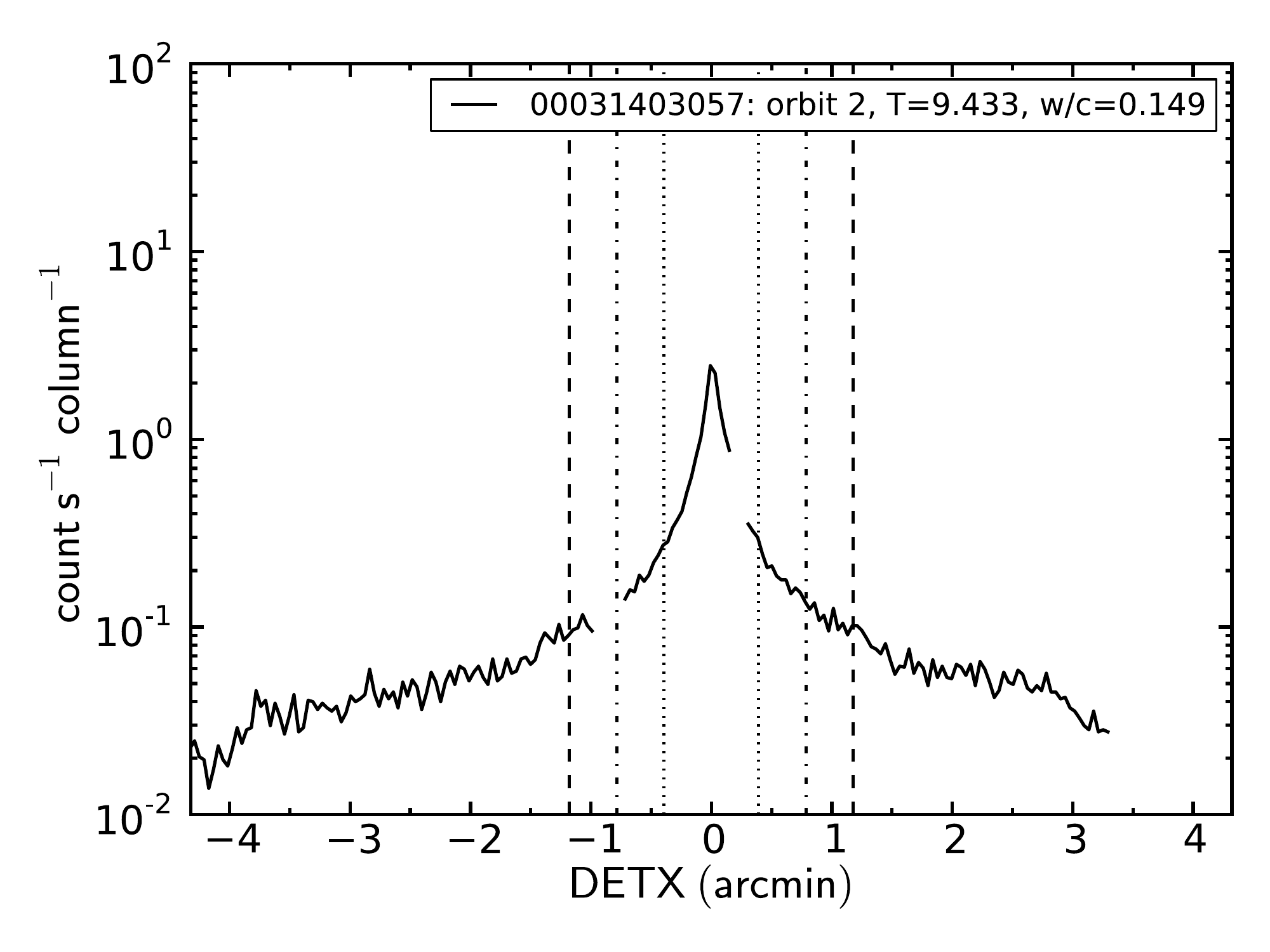}
\includegraphics[width=0.45\textwidth]{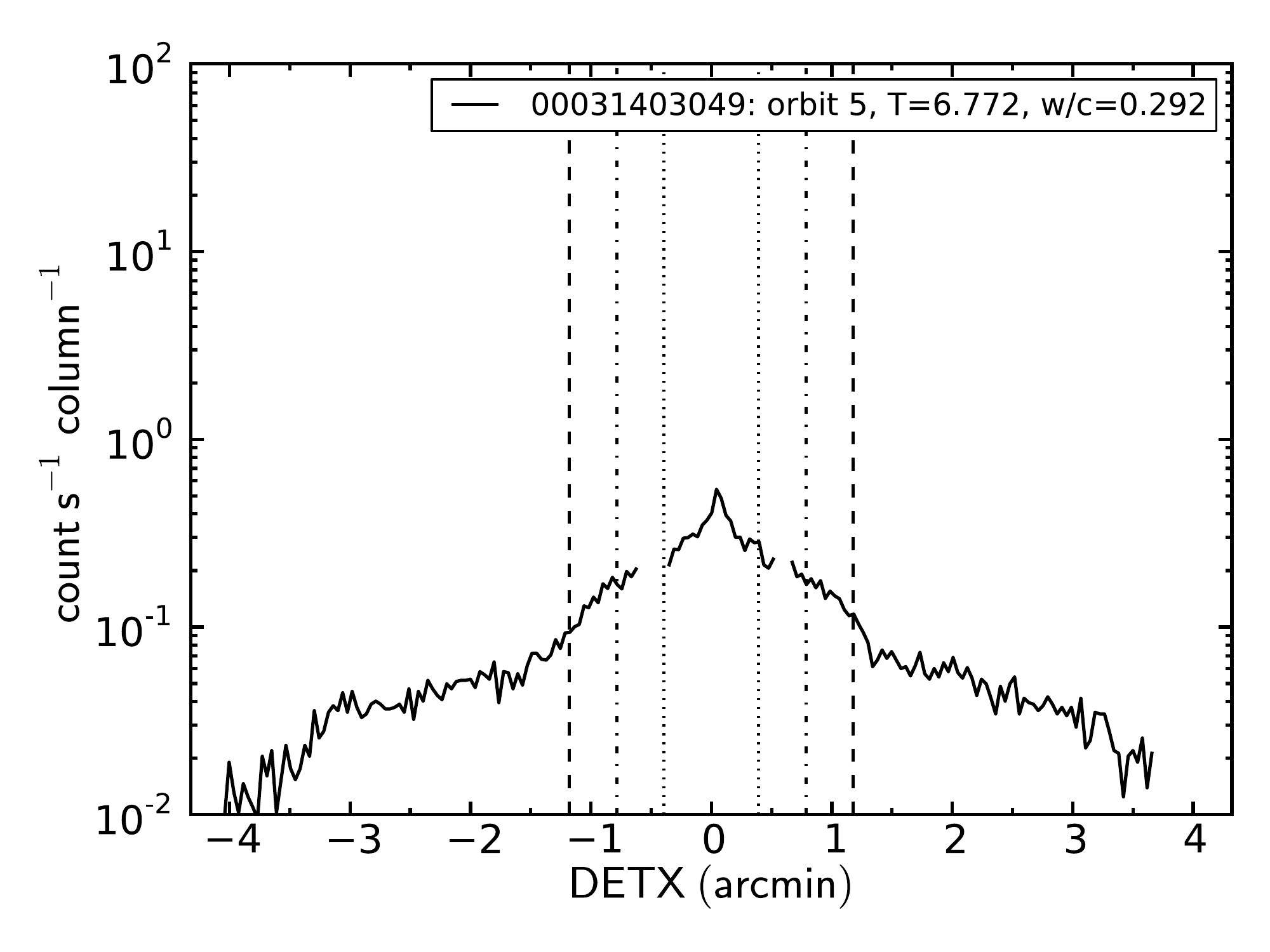} 
\includegraphics[width=0.45\textwidth]{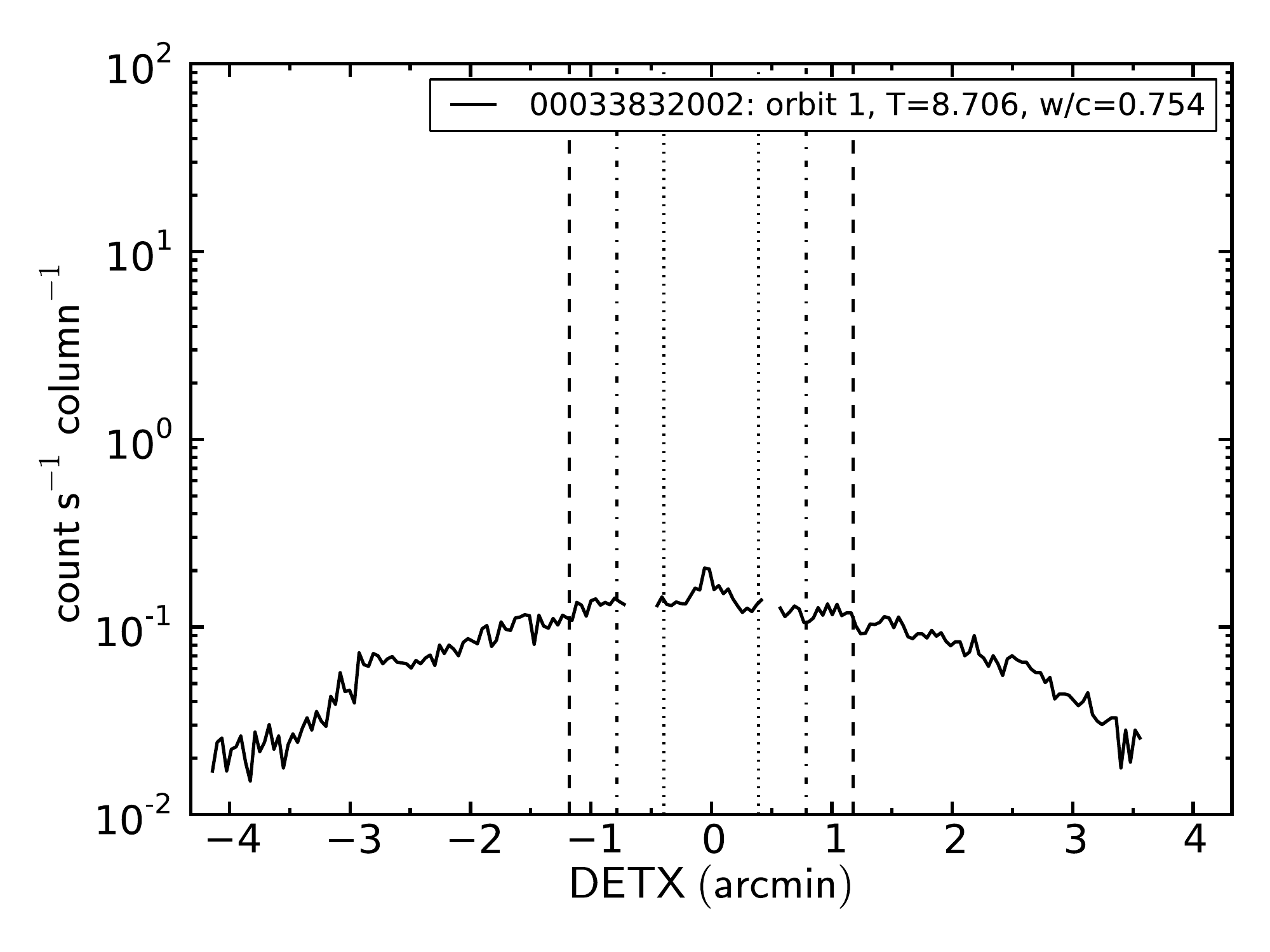}
\caption{Examples of the PSF 1D profiles extracted from observation 00031403049 snapshot 2, 00031403057 snapshot 2, 00031403049 snapshot 5,
00033832002 snapshot 1. The legend in each plot contains the Observation ID, the orbit (snapshot) number, the time since BAT trigger (MJD 57188.7727) and the wind-to-core ratio (see the text for details). The vertical lines mark the 10 (dotted), 20 (dotted-dashed) and 30 (dashed) pixels distance from the central source. }
\label{fig:profiles}
\end{figure*}

\begin{figure}
\includegraphics[width=0.48\textwidth]{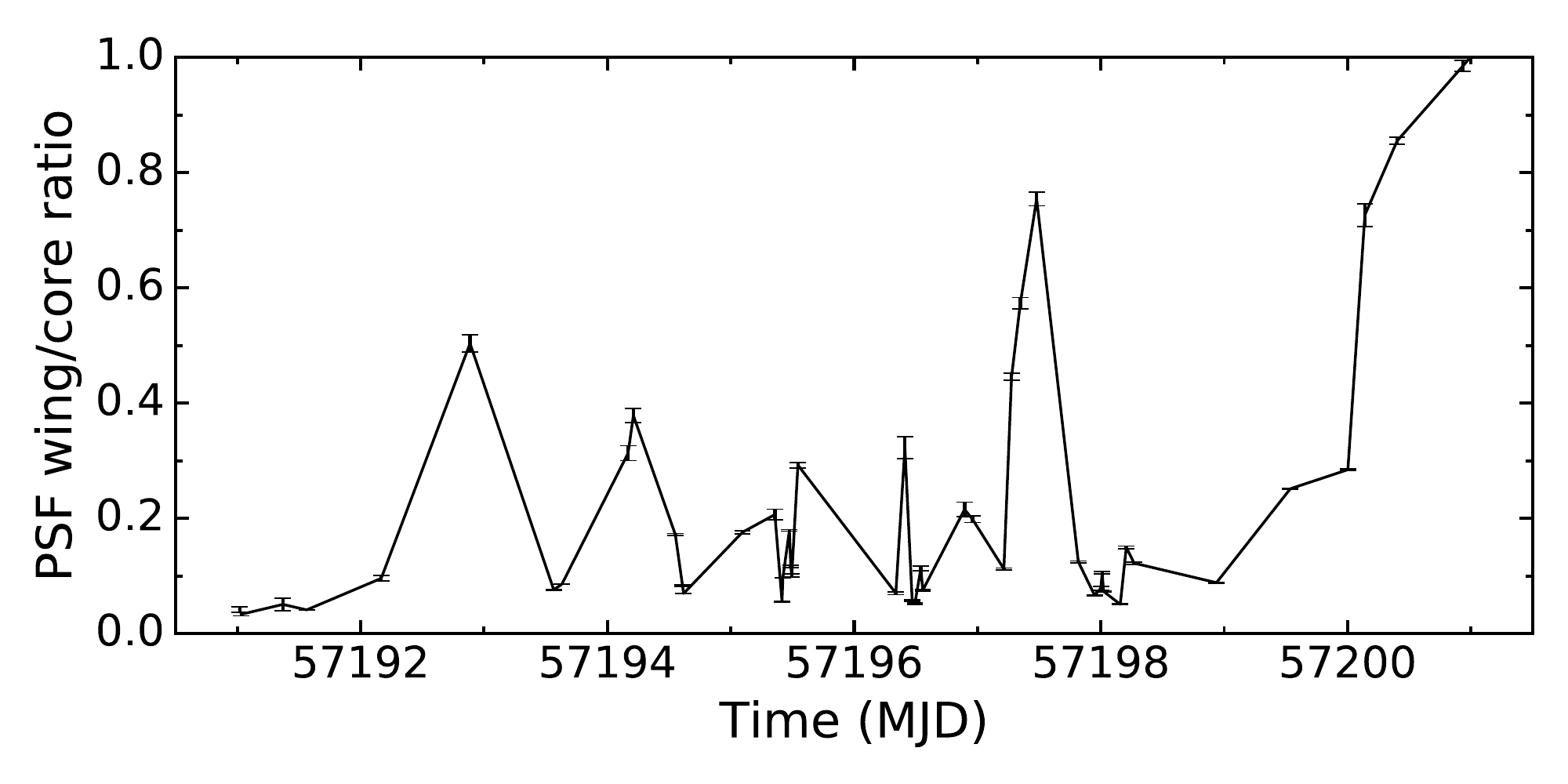}
\caption{PSF wing/core ratio, where the wing region is an annular region with radii 0.98 and 1.96 arcmin from the source position, respectively, while the core region is a circular region with radius 0.79 arcmin, centred on the source position. The wing and core count rates have been extracted in the 0.7--3.0 keV energy band in order to maximize the contribution of the halo with respect to that of the source. }
\label{fig:ratios}
\end{figure}


As mentioned in Sec. \ref{sec:analysis}, in this work we only consider WT mode observations, which did not allow us to produce a proper image of the source and of the dust scattering halo. However, we could still check if the dust scattering halo is present and whether it is affecting the source by extracting the 1-D WT detector profiles (see \citealt{Beardmore2016}), calculated over the restricted energy band of 0.7--3.0 keV, chosen to emphasize the soft dust scattering halo emission over that of the harder central source. 

In order to determine the role of the halo emission in the overall emission from the source, we first extracted the observed counts/s/column rate 1-D profiles for each snapshot of each observations considered in this work. Then, we extracted the source counts in a circular region with radius 20 pixels (0.786 arcmin, the \textit{core region}) and the halo counts in an annular region with inner and outer radius 25 and 50 pixels (the \textit{wing region}), respectively (see in Fig. \ref{fig:profiles} a few examples). As the CCD bad-columns can have a noticeable effect on the observed central source rates, we corrected them by interpolating the counts across the bad-columns to fill in the gaps. 
Then we constructed the ratio between the core and the wing count rates (the wing-to-core ratios, w/c), shown in Fig. \ref{fig:ratios}.  By comparing the 1-D psf profiles and the core-to-wing ratios, we determined that when the ratios exceeded 0.15, the dust scattering halo was likely affecting the source. For this reason, in snapshots with core-to-wing ratios larger than 0.15, we extracted the source spectrum in a circular region of only 10 pixels around the source, in order to minimize the halo contamination to the source emission. We note that the halo is typically dominating the field when the source is particularly faint, therefore extracting the spectrum in a small region is appropriate. 

We needed to adopt the above strategy since the halo spectrum cannot be systematically fitted together with each spectrum of our sample as in, e.g., \cite{Motta2017}. This is due to the limited S/N of our spectra, which in many cases would have not allowed us to constrain the parameters of a model including a component describing the halo as well.
However, in order to reduce to the minimum the possible contamination from the dust halo,  instead of extracting a background spectrum for each segment from the observation where it comes from, we extracted the background from a routine WT mode calibration observation of RXJ1856.4--3754 performed on 2015-Mar (exposure 17.8 ks, \citealt{Beardmore2016}). We selected an annular region with inner and outer radii fixed at 80 and 120 pixels (3.2 and 4.8 arcmin, respectively) away from the centre of the field (ra, dec = 284.17, -37.91).
We note, however, that the V404 Cyg count rate almost always exceeds 10~counts~s$^{-1}$ in our data set, thus subtracting the background counts has very little effect on the final spectrum (see http://www.swift.ac.uk/analysis/xrt/), since the final background spectrum count rate in the 0.6-10 keV energy band is 0.25 counts/s in a typical 20 pixel radius extraction region.

\section{Spectral modelling including reflection} \label{App:AppendixA}

\begin{figure*}
\centering
Parameter evolution from a partially-covered, absorbed power law convolved with a reflection component.
\includegraphics[width=0.94\textwidth]{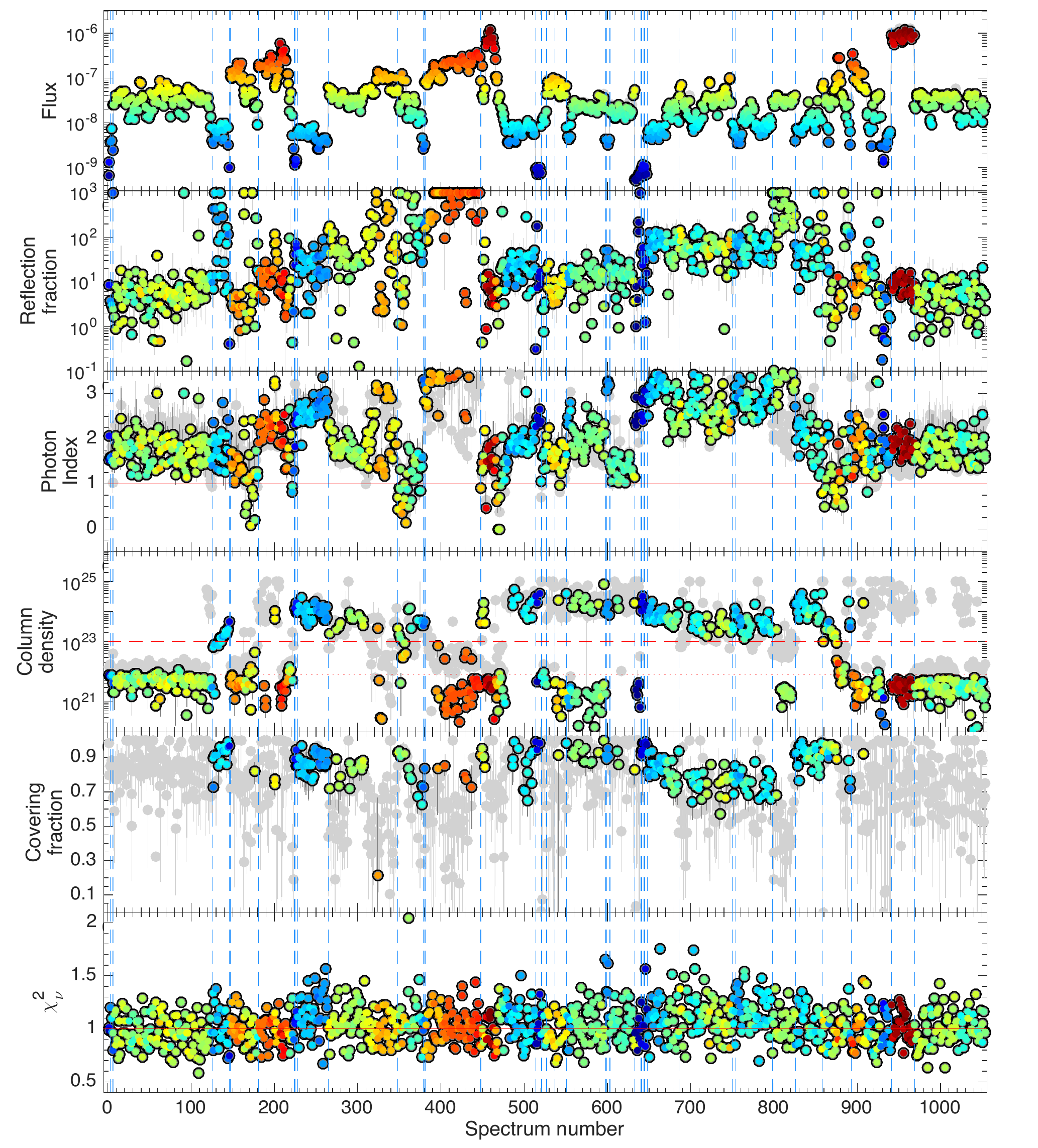}
\caption{Light-curve and parameter curves from our spectral fits. In this case, spectra were modelled with a partially-covered absorbed power law convoluted with a reflection kernel (\textsc{reflect}). All parameters are plotted as a function of spectrum number instead of time to facilitate parameter inspection. Points are colour coded as a function of Flux (in units of 10$^8$ erg/cm$^2$/s) corrected for the ISM absorption, shown in the first panel from the top. Other panels show: reflection fraction; spectral photon index; intrinsic column density; partial covering fraction; $\chi^2_\nu$. The solid red line marks photon index equal 1 in the photon index panel, column density equal 10$^{23}$cm$^{-2}$ (dashed) and column density equal to the IMS value (dotted) in the column density panel,  and $\chi^2_\nu$ equal 1 in the $\chi^2_\nu$ panels. 
As in Fig. \ref{fig:parameters}, we only plotted the points indicating covering fractions less than 1 in the partial covering fraction panel, since covering fraction equal 1 effectively means no covering fraction (uniform covering).  
The grey points show the parameter evolution obtained leaving the covering fraction and the local column density always free to vary, while the coloured points show the results of the fits to the spectra of our sample after applying the BIC test in order to determine whether the partial covering was statistically required (see Sec. \ref{sec:analysis} for details). A colour version of this Figure is available on-line.}
\label{fig:parameters_refl}
\end{figure*}

For the sake of comparison and completeness, we report in Fig. \ref{fig:parameters_refl} the results of the spectral fits obtained by fitting to the data an absorbed power law model convolved with a reflection component (\textsc{reflect}). The figure shows that in a few cases (e.g., in spectra 650-800) such spectral modelling seems to provide the best description of the data, even though the reflection component can rarely be constrained. The panels shown in Fig. \ref{fig:parameters_refl} are the same as in Fig. \ref{fig:parameters} save for the second panel from the top, where we show the evolution of the reflection factor.

\bigskip

\label{lastpage}
\end{document}